\documentclass{article}
\usepackage{amsmath}
\usepackage{subfigure}
\usepackage{float}
\usepackage{epstopdf}
\usepackage{amsfonts}
\usepackage{bm}
\usepackage{cases}
\usepackage{graphicx}
\usepackage{epsfig}  
\usepackage{calc}

\def\be#1\ee{\begin{equation}#1\end{equation}}
\def\ba#1#2\ea{\begin{eqnarray}#1#2\end{eqnarray}}

\newcommand{\bq}{\begin{equation}}
\newcommand{\eq}{\end{equation}}
\newcommand{\sgn}{\text{sgn}}

\usepackage{epsfig}  
\usepackage{calc}

\newtheorem{theorem}{Theorem}
\newtheorem{proposition}{Proposition}
\newtheorem{remark}{Remark}
\newtheorem{definition}{Definition}

\def\R{{\mathbb R}}

\usepackage{makeidx}         
\usepackage{graphicx}        
\usepackage{multicol}        
\usepackage[bottom]{footmisc}

\makeindex             

\begin{document}

\title{Kinetic models for socio-economic dynamics\\ of speculative markets}
\author{
Dario Maldarella\thanks{Department of Mathematics and CMCS, University of
Ferrara, Via Machiavelli 35 I-44100 Ferrara, Italy.
({\tt dario.maldarella@unife.it})}\and Lorenzo Pareschi\thanks{
Department of Mathematics and CMCS, University of
Ferrara, Via Machiavelli 35 I-44100 Ferrara, Italy.
({\tt lorenzo.pareschi@unife.it})}}
%
%
\maketitle

\begin{abstract}
In this paper we introduce a simple model for a financial market
characterized by a single stock or good and an interplay between
two different traders populations, chartists and fundamentalists,
which determine the price dynamic of the stock. The model has
been inspired by the microscopic Lux-Marchesi model \cite{ch4:LM}.
The introduction of kinetic equations permits to study the
asymptotic behavior of the investments and the price distributions
and to characterize the regimes of lognormal behavior and the
formation of power law tails.
\end{abstract}


\section{Introduction}

Most speculative markets at national and international level share
a number of stylized facts, like volatility clustering and fat
tails of returns, for which a satisfactory explanation is still
lacking in standard theories of financial markets
\cite{ch4:Pagan}. Such stylized facts are now almost universally
accepted among economists and physicists and it is now clear that
financial markets dynamics give rise to some kind of universal
scaling laws.

Showing similarities with scaling laws for other systems with many
interacting particles, a description of financial markets as
multi-agent interacting systems appeared to be a natural
consequence \cite{ch4:LLS1, ch4:LM, ch4:MS, ch4:SZSL, ch4:VT}.
This topic was pursued by quite a number of contributions
appearing in both the physics and economics literature in recent
years \cite{ch4:BM, ch4:AC1, ch4:CCM, ch4:YD, ch4:GGPS, ch4:IKR,
ch4:LLS, ch4:MS, ch4:PGS, ch4:VT}. This new research field borrows
several methods and tools from classical statistical mechanics,
where emerging complex behavior arises from relatively simple rules due to
the interaction of a large number of components \cite{NPT}.

Starting from the microscopic dynamics, kinetic models can be
derived with the tools of classical kinetic theory of fluids
\cite{ch4:BM, ch4:CPP, ch4:CPT, ch4:YD, ch4:DT, ch4:IKR, ch4:CMPP,
MP, ch4:MT, ch4:PT, ch4:SL}. In contrast with microscopic dynamics,
where behavior often can be studied only empirically through
computer simulations, kinetic models based on PDEs allow us to
derive analytically general information on the model and its
asymptotic behavior.


In this paper we introduce a simple Boltzmann-like model for a
speculative market characterized by a single stock and
a socio-economical interplay between two different types of traders,
chartists and fundamentalists. The model is strictly related to
the microscopic Lux-Marchesi model \cite{ch4:LM} and to kinetic models of opinion
formation recently introduced in \cite{ch4:TG}. In addition, we
take into account some psychological and behavioral components of
the agents, like the way they interact each other and perceive
the risk, which may produce non rational behaviors. This is done
by means of a suitable ``value function'' in agreement with the
Prospect Theory by Kahneman and Tversky \cite{ch4:KT, ch4:KT1}.
As we will show people systematically overreacting produces
substantial instabilities in the stock market.

In an earlier paper \cite{ch4:CPP} a similar approach has been
used considering a single population of investors interacting in
the stock market on the basis of the microscopic Levy-Levy-Solomon
model \cite{ch4:LLS}. The emergence of a lognormal behavior for
the wealth distribution of the agents has been shown.
Though the theoretical set-up of the analysis is close in certain respects
to that of \cite{ch4:CPP}, the structure of the model is rather
different. Namely, the description of individual behavior follows an
opinion formation dynamic strictly connected with the price trend.
In this way, the heterogeneity among agents as well as their social
interactions will be taken into account which both are key elements affecting
the outcome of the overall market dynamics.

Following the analysis developed in
\cite{ch4:CPT, ch4:TG}, we shall prove that the Boltzmann model
converges in a suitable asymptotic limit towards
convection-diffusion equations of Fokker-Planck type. Other
Fokker-Planck equations were obtained using different approaches
in \cite{ch4:BM, ch4:SMBSR, ch4:So}. This permits to study the
asymptotic behavior of the investments and the price distributions
and to characterize the regimes of lognormal behavior
and the ones with power law tails. The main finding of the present paper
is that the presence of heterogeneous strategies, both fundamentalists
and chartists, is essential to achieve basic stylized fact like the presence
of fat tails.

The rest of the paper is organized as follows. In Section 2 we introduce the
Boltzmann kinetic model for the interacting chartists and the price evolution.
Details of the strategy exchange between chartists
and fundamentalists are also presented here. A characterization of the admissible
equilibrium states of the resulting system is then reported.
Next, in Section 3, with the aim to study the asymptotic behavior of the chartists
and price distributions, we introduce simpler Fokker-Planck approximations of the
Boltzmann system and give explicit expressions of the long time behavior. The mathematical
details of the derivation of such Fokker-Planck models are reported in separate appendices
at the end of the manuscript.
Numerical results which confirm the theoretical analysis are given in Section 4 and some concluding
remarks are discussed in the last section.

\section{A kinetic model for multiple agents interactions}
\label{ch4:sec:MP} We describe a simple financial market
characterized by a single stock or good and an interplay between
two different traders populations, chartists and fundamentalists,
which determine the price dynamic of such stock (good). The
aim is to introduce a kinetic description both for the behavior
of the microscopic agents and for the price, and then to exploit
the tools given by kinetic theory to get more insight about the
way the microscopic dynamic of each trading agent
can influence the evolution of the price, and be responsible of
the appearance of 'stylized' fact like 'fat tails' and 'lognormal'
behavior.

\subsection{Kinetic setting} Similarly to Lux and Marchesi model
\cite{ch4:LM}, the starting point is a population of two different
kind of traders, chartists and fundamentalists. Chartists are
characterized by their number density $\rho_C$ and the investment
propensity (or opinion index) $y$ of a single agent whereas
fundamentalists appear only through their number density $\rho_F$.
The value $\rho=\rho_F+\rho_C$ is invariant in time so that the
total number of agents remains constant. In the sequel we will
assume for simplicity $\rho=1$.

\paragraph{Dynamic of investment propensity among chartists.}

Let us define $f(y,t)$, $y\in[-1,1]$, the distribution function of
chartists with investment propensity $y$ at time $t$. Positive values of $y$
represent buyers, negative values characterize sellers and close
to $y=0$ we have undecided agents. Clearly
\[
\rho_C(t)=\int_{-1}^1 f(y,t)\,dy.
\]
Moreover we define the mean investment propensity \be Y(t)=
\frac1{\rho_C(t)}\int_{-1}^1 f(y,t)y\,dy.\ee For a given price
$S(t)$ and price derivative $\dot{S}(t)=dS(t)/dt$ the microscopic
dynamic of the investment propensity of chartists is characterized
by the following binary interactions $(y,y_*)\to(y',y'_*)$ with
\begin{eqnarray}
\nonumber
y'&=&(1-\alpha_1 H(y)-\alpha_2)y+\alpha_1 H(y){y}_*+\alpha_2\Phi\left(\frac{\dot{S}(t)}{S(t)}\right)+
D(y)\eta,\\[-.2cm]
\\
\label{eq:bin}
\nonumber y_*'&=&(1-\alpha_1
H(y_*)-\alpha_2)y_*+\alpha_1
H(y_*){y}+\alpha_2\Phi\left(\frac{\dot{S}(t)}{S(t)}\right)+D(y_*)\eta_*.
\end{eqnarray}
Here
$\alpha_1\in [0,1]$ and $\alpha_2 \in [0,1]$, with $\alpha_1 +
\alpha_2 \leq 1$, measure the importance the individuals place on
others opinions and actual price trend in forming expectations
about future price changes. The random variables $\eta$ and
$\eta_*$ are assumed distributed accordingly to $\Theta(\eta)$
with zero mean and variance $\sigma^2$ and measure individual
deviations from the average behavior. The function $H(y)\in [0,1]$
is taken symmetric on the interval $I$, and characterize the
herding behavior, whereas $D(y)$ defines the diffusive behavior,
and will be also taken symmetric on $I$. Simple examples of
{herding function} and {diffusion function} are given by
\[
H(y)=a+b(1-|y|),\qquad D(y)=(1-y^2)^{\gamma},
\]
with $0\leq a+b\leq 1$, $a\geq 0, b>0$, $\gamma > 0$ (see Figure \ref{fig1}). Other
choices are of course possible, note that in order to preserve the
bounds for $y$ it is essential that $D(y)$ vanishes in $y=\pm 1$.
Both functions take into account that extremal positions suffer
less herding and fluctuations. For $b=0$, $H(y)$ is constant and
no herding effect is present and the mean investment propensity is
preserved when the market influence is neglected ($\alpha_2=0$)
 as
in classical opinion models a model ( see \cite{ch4:TG} at the reference therein).

\begin{figure}[H]
\begin{center}
\includegraphics[scale=0.35]{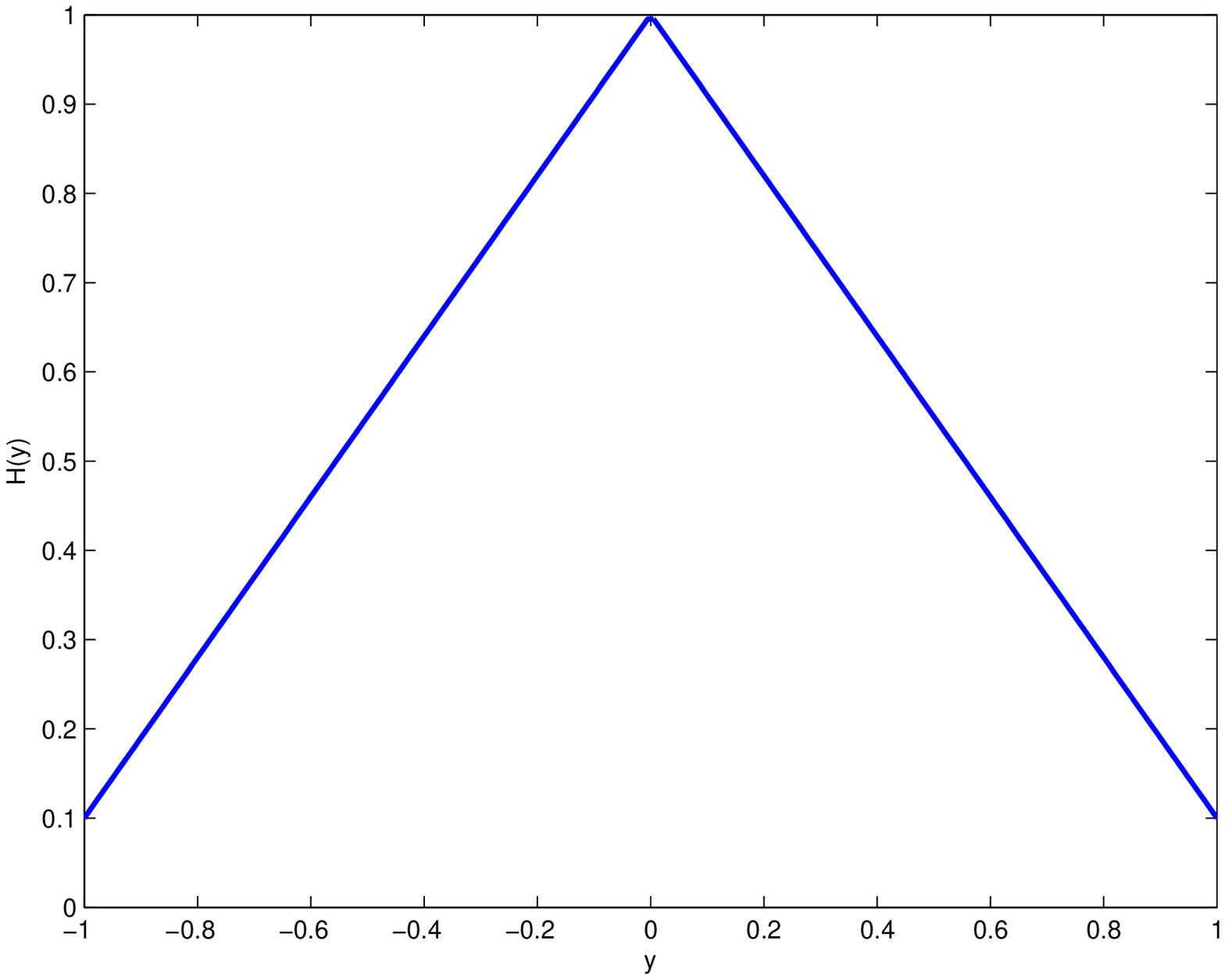}\,
\includegraphics[scale=0.35]{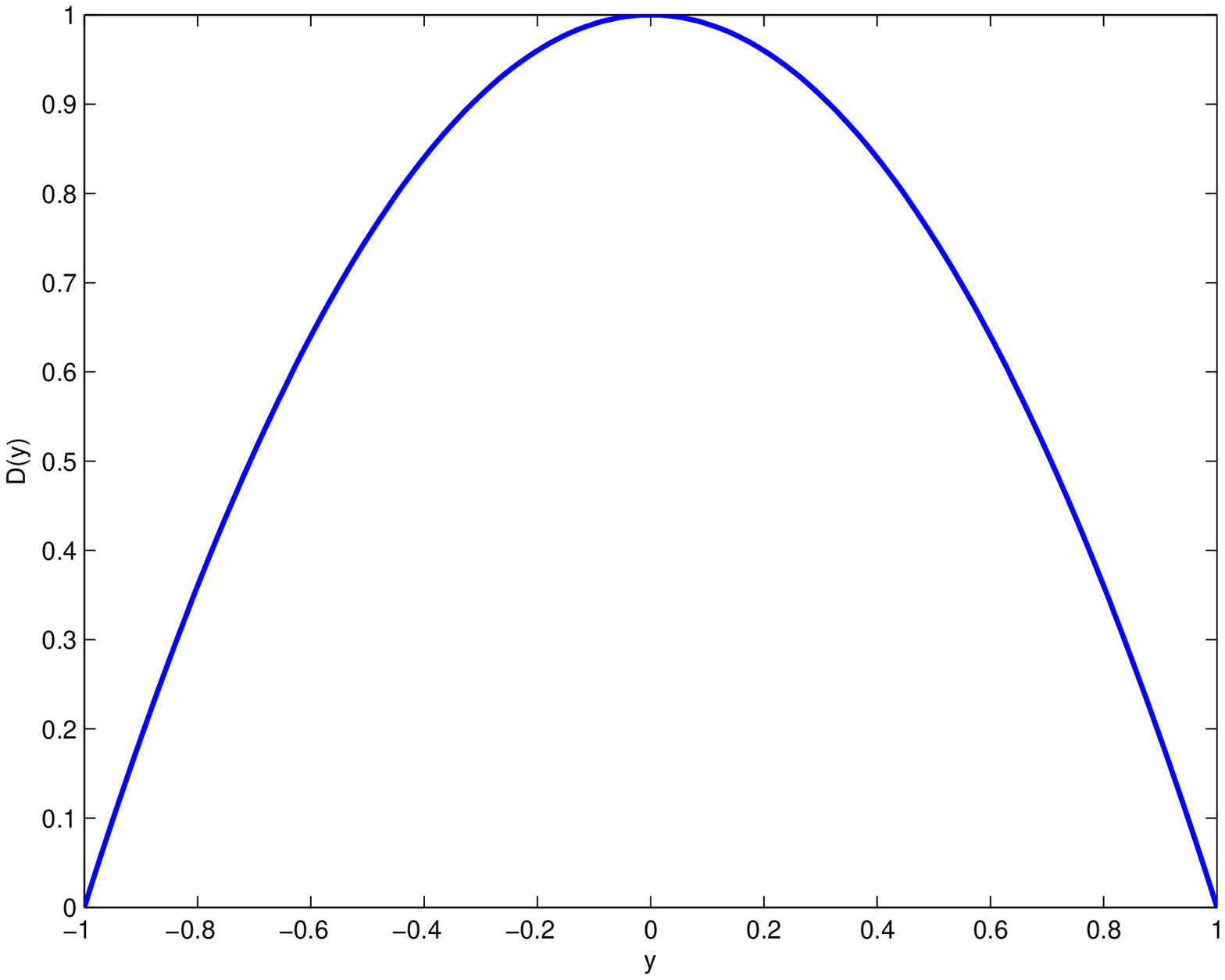}
\end{center}
\caption{Typical examples of herding function $H(y)$ (left) and
diffusion function $D(y)$ (right).}
\end{figure}\label{fig1}

%

A remarkable feature of the above relations is the presence of the
normalized value function $\Phi(\dot{S}(t)/{S(t)})$ in $[-1,1]$ in
the sense of Kahneman and Tversky \cite{ch4:KT, ch4:KT1} that
models the reaction of individuals towards potential gain and
losses in the market \cite{ch4:KT}. This permits to introduce
behavioral aspects in the market dynamic and to take into account the
influence of psychology on the behavior of financial
practitioners.

The value function is defined on deviations from a reference
point, which is usually assumed  equal to zero (but it can be
considered also positive or negative), and is normally concave for
gains (implying risk aversion), commonly convex for losses (risk
seeking) and is generally steeper for losses than for gains (loss
aversion)(see Figure \ref{fig2}).
\begin{figure}[H]
\begin{center}
\includegraphics[scale=.35]{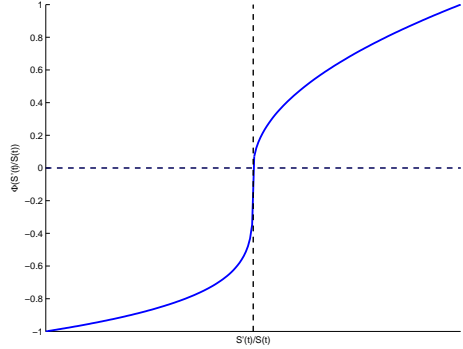}
\end{center}
\caption{An example of value function $\Phi(\dot{S}(t)/S(t))$.}
\end{figure}\label{fig2}

Let us ignore for the moment the price evolution. The above binary
interaction gives the following kinetic equation for the time
evolution of chartists
\begin{equation}\label{Boltzmanchart}
\frac{\partial f}{\partial t} =
Q(f,f),
\end{equation}
where  for any test function $\varphi$ the interaction operator
$Q$ can be conveniently written in weak form as \be
\int_{-1}^{1}Q\varphi(y)\,dy=\int_{[-1,1]^2}\int_{\R^2}
B(y,y_*,\eta,\eta_*)
f(y)f(y_*)(\varphi(y')-\varphi(y))d\eta\,d\eta_*\,dy_*\,dy \ee
with the transition rate has given by

\[
B(y,y_*,\eta,\eta_*)=\Theta(\eta)\Theta(\eta_*)\chi(|y'|\leq
1)\chi(|y_*'|\leq 1),
\]
being $\chi(\cdot)$ the indicator function.
Note that the mass density of chartists $\rho_C(t)$ is an
invariant for the interaction, ($\varphi\equiv 1$).\\
It is worth to observe that for a given $D_C(y)$ a suitable choice of the support of the random variable $\eta$,
avoids the dependence
of the collisional kernel $B(y,y_{*})$ on the variables $y$,$y^{*}$. \\
As an example, if we take $D(y)=1-y^2$ we have
\begin{eqnarray*}
 y'&=& (1-\alpha_1H(y)-\alpha_2)y +\alpha_1H(y)y_{*}+\alpha_2\Phi\left(\frac{\dot{S(t)}}{S(t)}\right) + (1-y^2)\eta\\
&\leq&
(1-\alpha_1H(y)-\alpha_2)y +\alpha_1H(y) + \alpha_2 + (1-y^2)\eta.
\end{eqnarray*}
 Then to have $y'\leq1$ for any $y\in[-1,1]$, we have to chose $\eta$ such that
\begin{eqnarray*}
 (1-y^2)\eta\leq(1-\alpha_1-\alpha_2)(1-y)
\end{eqnarray*}
which gives
$$\eta \leq \frac{1}{2}(1-\alpha_1-\alpha_2).$$
Analogously we can ensure $y'\geq -1$ , thus it is enough to take
$$\eta\in[-\frac{1}{2}(1-\alpha_1-\alpha_2),\frac{1}{2}(1-\alpha_1-\alpha_2)].$$
For this reason, in the rest of the paper, we will consider only kernel of ``maxwellian type''
%

$$B(y,y_{*},\eta,\eta_*) = \Theta(\eta)\Theta(\eta_{*}).$$

\paragraph{Strategy exchange chartists-fundamentalists.}

In addition to the change of investment propensity due to a
balance between herding behavior and the price followers nature of
chartists, the model includes the possibility that an agent changes
its strategy from chartist to fundamentalists and viceversa.

Agents meet individual from the other group, compare excess
profits from both strategies and with a probability depending on
the pay-off differential switch to the more successful strategy.
When a chartist and a fundamentalist meet they characterize the
success of a given strategy trough the profits earned by comparing
\begin{equation}\label{profit}
X_{C}(y,t) =\psi(y)\left(\frac{\dot{S}(t)/\mu+D}{S(t)} -r\right), \quad X_{F}(t)= k\frac{|S_{F}-S(t)|}{S(t)}.
\end{equation}
Here $\psi(y)\in [-1,1]$ has the same sign of $y$ and takes into
account the change of sign in the profits accordingly to the
actual behavior of the agent in the market which rely on his
investment propensity $y$. The simplest choice is $\psi(y)=\sgn(y)$.

The value $D$ is the nominal dividend and $r$ the average real
return of the market, such that $r={D}/{S_{F}}$, i.e. evaluated at
its fundamental value $S_{F}$ in a state of stable price
$\dot{S}=0$ the asset yield the same returns of other investments,
 or equivalently $X_{C}=X_{F}=0$. The discount factor $k<1$ is justified by the observation that $X_{F}$ is an expected gain realized
only after reversal to the fundamental value.
Finally  $\mu>0$ measures the frequency of the exchange rates.

A chartist characterized by an investment propensity $y$  and a
fundamentalist meet each other, and after comparing their
strategies, they exchange strategies with a rate given by a
suitable monotone function $B_{FC}(\cdot)\geq 0$. More precisely a
chartist switch to fundamentalist with a rate $B_{FC}(X_F-X_C)$ and a
fundamentalist switch to chartist at a rate $B_{FC}(X_C-X_F)$.

For chartists we define the following linear strategy exchange
operator
$$ Q_{FC}(f) =  \mu \rho_{F}(t)f(y)(B_{FC}(X_{C}-X_{F}) - B_{FC}(X_{F}-X_{C}))$$
where $\mu>0$ measures the frequency of the exchange rates.

Taking into account such strategy exchanges we have the
chartists-fundamentalists model
\begin{equation}
\left\{
\begin{array}{lcl}
 \displaystyle \frac{\partial f}{\partial t} = Q(f,f) + \mu \rho_{F}(t)f(y)(B_{FC}(X_{C}-X_{F}) - B_{FC}(X_{F}-X_{C}))\\ [+0.25cm]
\displaystyle \frac{\partial \rho_F}{\partial t} = \mu
\rho_{F}(t)\int_{-1}^1 f(y)(B_{FC}(X_{F}-X_{C}) -
B_{FC}(X_{C}-X_{F}))\,dy.
\end{array}
\right. \label{ch4:eq:fc}
\end{equation}
It is immediate to verify that the total number density
$\rho_C+\rho_F$ is conserved in time.

\paragraph{Price evolution.}
Finally we introduce the probability density $V(s,t)$ of a given
price $s$ at time $t$. The effective market price $S(t)$ is defined as
the mean value
\be
S(t)=\int_0^\infty V(s,t) s\,ds.
\ee
Following Lux and Marchesi \cite{ch4:LM} the microscopic dynamic
of the price is given by \be s' = s + \beta (\rho_{C}t_{C}Y(t)s +
\rho_{F}\gamma(S_{F}-s)) + \eta s \ee where the parameters
$\beta$, represent the price speed evaluation, $\eta$ is a random
variable with zero mean and variance $\zeta^2$, distributed
accordingly to $\Psi(\eta)$. In the above relation chartists
either buy or sell the same number $t_C$ of units and $\gamma$ is
the reaction strength of fundamentalists to deviations from the
fundamental value.

Thus the chartists-fundamentalists system of equations
(\ref{ch4:eq:fc}) is complemented with the equation for the price
distribution \be \frac{\partial V}{\partial t} = L(V),
\label{ch4:eq:price}\ee where the operator $L$, is linear, and  in weak form it
reads
\begin{equation}\label{Priceoperator}
\int_{0}^{\infty}L(V)(s)\varphi(s)\,ds=\int_{0}^{\infty}\int_{\R}
b(s,\eta) V(s)(\varphi(s')-\varphi(s))d\eta\,ds
\end{equation}
with the transition rate $b(s,\eta)=\Psi(\eta)\chi(s'\geq 0). $\\
As before, a suitable choice of the
domain for the support of variable $\eta$ ensures $s'\geq 0$.
Assuming $$\eta \in [-1+\beta(\rho_{C}T_{C}+\rho_F\gamma),1-\beta(\rho_{C}T_{C}+\rho_F\gamma)],\quad \beta(\rho_{C}T_{C}+\rho_F\gamma) < 1,$$
permits to express the transition rate in the simpler form
 $$b(s,\eta)=\Psi(\eta). $$

Note that the expected value for the stock price satisfies the
same differential equation 
as in\cite{ch4:Lux,
ch4:LM} \be \frac{d S(t)}{dt} = \beta\rho_C t_C Y(t)S(t)
+\beta\rho_F \gamma(S_F - S(t)). \ee

\paragraph{Booms, crashes and macroscopic stationary states.}

In order to study the macroscopic steady states and relate them to the value function
$\Phi$ let us start by observing that the equilibrium states for the price
satisfy
$$\rho_C t_C Y S +\rho_F \gamma(S_F - S)=0$$
 and thus fall in one
of the following categories
\begin{itemize}
\item[(i)]$\rho_F\neq 0,\quad S=\frac{\rho_F\gamma S_F}{\rho_F\gamma-\rho_{C}t_{C}Y},\ \quad \rho_F\gamma S_F-\rho_C t_C Y\geq 0$.
\item[(ii)] $\quad \rho_F=0, \quad Y=0,\quad S\,{\rm arbitrary},$
\item[(iii)]$\quad \rho_F=0, \quad S=0,\quad Y\,{\rm arbitrary}.$
\end{itemize}
At equilibrium we require $\rho_F$, $\rho_C$ and $Y$ to be
constants. In order for the number densities to be constants we
require $Q_{FC}=0$. For $\rho_F\neq 0$ and $\rho_C\neq 0$, thanks
to monotonicity of $B_{FC}$, we have $X_C=X_F$ or equivalently
$S=S_F$. Note that $Q_{FC}$ vanishes also when $\rho_F=0$ or
$\rho_C=0$. These considerations reduce the set of possible
equilibrium configurations to
\begin{itemize}
\item[(i)]
$ \quad \rho_F\neq 0,\quad S=S_F,\quad Y=0,$
\item[(ii)] $\quad \rho_F=0, \quad Y=0,\quad S\,{\rm arbitrary},$
\item[(iii)]$\quad \rho_F=0, \quad S=0,\quad Y\,{\rm arbitrary}.$
\end{itemize}
Finally we consider the requirements for $Y$ to be constant. In
the case $Q_{FC}=0$ the first moment equation reads
\begin{eqnarray*}
\frac{d}{dt}Y(t) = &-&\alpha_{1}\int_{-1}^1H(y)yf(y)dy -
\alpha_{2}\rho_{C}Y(t)\\ &+&\alpha_{1}Y(t)\int_{-1}^1H(y)f(y)dy
      + \alpha_{2}\rho_{C}\Phi
      \left(\frac{\dot{S}(t)}{S(t)}\right),
\end{eqnarray*}
which gives the steady state condition \[
-\alpha_{1}\int_{-1}^1H(y)yf(y)dy - \alpha_{2}\rho_{C}Y
+\alpha_{1}Y\int_{-1}^1H(y)f(y)dy
      + \alpha_{2}\rho_{C}\Phi
      \left(\frac{\dot{S}(t)}{S(t)}\right)=0.
\label{ch4:eq:Yeq} \]
This gives a constraint for the value function $\Phi$, precisely

\begin{eqnarray*}
 \alpha_{2}\rho_{C}\Phi\left(\frac{\dot{S}(t)}{S(t)}\right)=\alpha_{1}\int_{-1}^{1}H(y)yf(y)dy + \alpha_{2}\rho_{C}Y
- \alpha_{1}Y\int_{-1}^{1}H(y)f(y)dy
\end{eqnarray*}
which in the simple case
of $H$ constant reduces to
\[
\alpha_2\rho_C
\left(\Phi\left(\frac{\dot{S}(t)}{S(t)}\right)-Y\right)=0.
\]
Now using the fact that \[\frac{\dot{S}(t)}{S(t)} = \beta\rho_C
t_C Y(t) +\beta\rho_F \gamma\frac{(S_F - S(t))}{S(t)},\] we can state

\begin{proposition}
The system of equations (\ref{ch4:eq:fc}) 
in the case of $H$ constant admits the following possible equilibrium
configurations
\begin{itemize}
\item[(i)] $\quad \rho_F\neq 0,\quad S=S_F,\quad Y=0,\quad \Phi(0)=0$,
\item[(ii)] $\quad \rho_F=0,\quad Y=0, \quad \Phi(0)=0, \quad
S$\,{arbitrary},
\item[(iii)] $\quad \rho_F=0,\quad Y=Y_*$,\, with\, $Y_*=\Phi(\beta t_C Y_*), \quad S=0$.
\end{itemize}
\label{ch4:prop:MP}
\end{proposition}

Note that if the reference point for the value function
$\Phi(0)\neq 0$ configuration $(i)$ and $(ii)$ are not possible
for a constant $H$. This is in good agreement with the fact that
an emotional perception of the market from the chartists acts as a
source of instability for the market itself. In contrast
configuration $(iii)$, corresponding to a market crash, can be
achieved also for $\Phi(0)\neq 0$. The existence of a unique fixed
point $Y_*$ has to be guaranteed by the choice of $\Phi$, $\beta$
and $t_C$. Of course if the reference point is set to zero,
$\Phi(0)=0$, we have $Y_*=0$. It is easy to verify that these
possible equilibrium configurations include the ones in the
original Lux-Marchesi model \cite{ch4:Lux}.

In addition to the above equilibrium configurations the model
admits several other possible asymptotic behavior in the form of
booms and cycles. Some of the fundamental features of the model
are summarized in the following.
\begin{remark}~\\
\begin{itemize}
\item Chartists alone ($\rho_F=0, \rho_C=1$) influence the price through their mean
propensity to invest $Y(t)$ and at the same time the price trend
influences their mean propensity through the value function
$\Phi(\dot{S}(t)/S(t))$, since $\dot{S}(t)/S(t)=\beta Y(t)t_C$.
Thus,  except for the particular shape of the value function, if
the mean propensity is initially (sufficiently) positive then it
will continue to grow together with the price and the opposite
occurs if it is initially (sufficiently) negative.

The market goes towards a {boom} (exponential grow of the price)
or a {crash} (exponential decay of the price) with
\[
S(0)e^{-\beta t_C}\leq S(t) \leq S(0)e^{\beta t_C},
\]
and agents tend to concentrate in $y=1$ and $y=-1$ respectively
depending on the choices of $H$ and $\Phi$. This is in good
agreement with the price followers nature of chartists.
\item Fundamentalists alone ($\rho_F=1, \rho_C=0$) influence the price through their expectation
of the fundamental price. So their effect is to drive the price
towards the fundamental price. For a constant fundamental price
$S_F$ the equilibrium state reached is characterized by $S=S_F$
and the trend is exponential.

\item The presence of fundamentalists acts in contrast to the chartists
pressure towards market booms or crashes. If their number is large
enough they are capable to drive the price towards the fundamental
value otherwise the chartists dynamic may dominate. In addition to
booms and crashes, we have now the possibility of price
cycles/oscillations around the fundamental value.

\end{itemize}
\end{remark}

\section{Fokker-Planck approximations and asymptotic behavior}

Now we consider what happens at the kinetic scale. Due to the
extreme difficulty to get detailed information on the asymptotic
behavior of the kinetic coupled system, we will recover for both
distribution functions $f$, and $V$, simplified Fokker-Planck
models which preserve the main features of the original kinetic
model. To keep notations simple, since we are mostly interested in
the study of the equilibrium states we ignore the presence of the
terms describing the change of strategy. However they can be
easily included in the scaling described below. 

For this purpose we introduce a time scaling parameter $\xi$ and
define \be
\tau = \xi t,\quad \tilde f(y,\tau) =  f(y, t),\quad
\tilde V(s,\tau) = V(s, t). \label{eq:sc1} \ee
 To preserve the
chartists dynamic in the limit, we must require that \be
\lim_{\alpha_1,\xi\rightarrow
0}\frac{\alpha_{1}}{\xi}=\tilde{\alpha_1},\quad
\lim_{\alpha_2,\xi\rightarrow
0}\frac{\alpha_{2}}{\xi}=\tilde{\alpha_2}, \quad
\lim_{\sigma,\xi\rightarrow 0}\frac{\sigma^{2}}{\xi}= \lambda,
\label{eq:sc2} \ee where $\lambda$ is a positive constant.

Similarly for the price dynamic, we assume
\be\lim_{\beta,\xi\rightarrow 0}\frac{\beta}{\xi}=
\tilde{\beta},\quad \lim_{\zeta,\xi\rightarrow
0}\frac{\zeta^2}{\xi}= \nu. \label{eq:sc3}
\ee

Performing similar computations as in \cite{ch4:CPP} (see Appendix
A and B for details) we recover the following Fokker-Planck
system 
\begin{subnumcases}{}
\displaystyle\frac{\partial \tilde{f}}{\partial \tau}
+\frac{\partial}{\partial y}\left[\rho_C \left(\tilde{\alpha_{1}}
H(y)(\tilde{Y}-y) + \tilde{\alpha_{2}} \left(\tilde\Phi
-y\right)\right)\tilde{f}\right] \displaystyle =
\frac{\lambda\rho_{C}}{2}\frac{\partial^{2}}{\partial
y^{2}}({D}^{2}(y)\tilde{f}), &\label{8a}\\
\displaystyle\frac{\partial}{\partial \tau}\tilde{V} +
\frac{\partial}{\partial
s}\left[\tilde{\beta}\left(\rho_{C}\tilde{Y} t_{C}s +
\rho_F\gamma(S_F-s)\right)\tilde{V}\right] =
\frac{\nu}{2}\frac{\partial^{2}}{\partial s^{2}}\left(
s^{2}\tilde{V}\right), &\label{8b}
\end{subnumcases}
where we used the shorthand $\tilde\Phi$ for
$\Phi\left({\dot{\tilde{S(\tau)}}}/{\tilde{S(\tau)}}\right)$.

For notation simplicity in the sequel we will omit the tildes in
the variables $f$, $V$, $Y$ and $S$.

If we now take $D(y) = 1 - y^{2}$, and  $H(y) = 1 $ we can
compute explicitly the equilibrium state for chartists with a
constant mean investment propensity ${Y}=Y_*$ as

\begin{equation}\label{ch4:chartistfokkplank2}
\begin{array}{lcl}
{f}^{\infty}(y) &=&\displaystyle
C_{0}(1+y)^{-2+Y_*\frac{(\tilde{\alpha_{1}}+\tilde{\alpha_{2}})}{2\lambda}}
(1-y)^{-2-Y_*\frac{(\tilde{\alpha_{1}}+\tilde{\alpha_{2}})}{2\lambda}}
\exp\left(\displaystyle -\frac{(1-Y_*
y)(\tilde{\alpha_{1}}+\tilde{\alpha_{2}})}{\lambda(1-y^{2})}\right)
\end{array}
\end{equation}
where
$C_0=C_0({Y_*,\lambda/(\tilde{\alpha_{1}}+\tilde{\alpha_{2}})})$
is such that the mass of ${f}^{\infty}$ is equal to $\rho_C$.
Other choices of the diffusion function originate different steady
states (see \cite{ch4:TG}).

\begin{figure}[t]
\begin{center}
{\includegraphics[scale=.40]{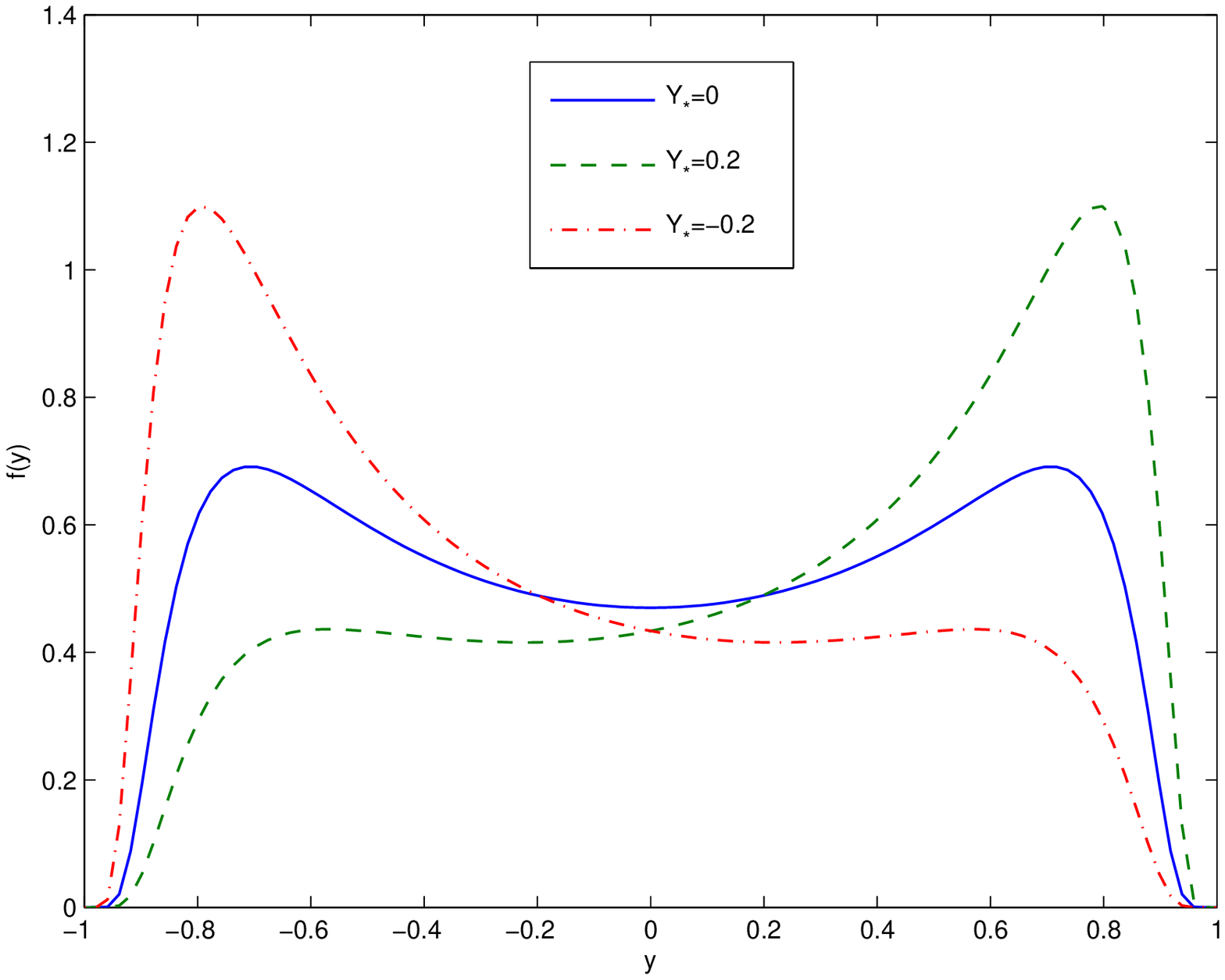}}\,
{\includegraphics[scale=.40]{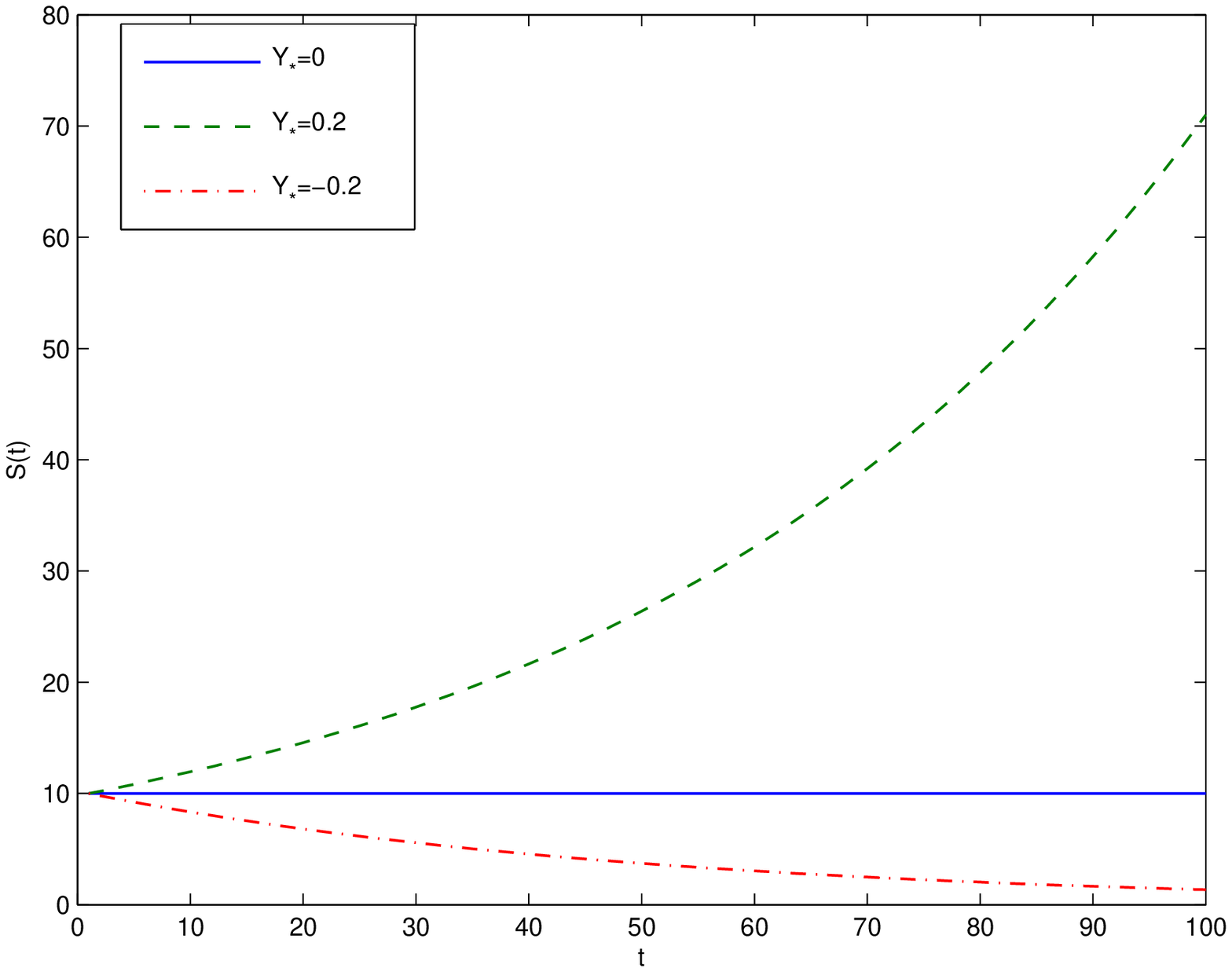}}
\end{center}
\caption{Equilibrium distribution function of the chartist
investment propensity for different values of $Y_*=0,0.2,-0.2$
(left) and corresponding behavior of the price $S$ (right). Exact
solutions with $\rho_C=1$, $\beta=0.1$, $t_C=1$,
$\lambda/(\tilde{\alpha_{1}}+\tilde{\alpha_{2}})=1$ and
$f(y,0)=f^{\infty}(y)$.} \label{ch4:fg:eqf}
\end{figure}

Observe that, in the case $Y_*\neq0 $, the distribution is not
symmetric and in the chartist population a predominant behavior
arise. Otherwise when the reference point of the value function is
set to zero we have a symmetric distribution with two peaks and
mean value zero, and the macroscopic state of indecision is given,
microscopically, by a polarization of the chartist population
among two opposite kind of behaviors (see Figure
\ref{ch4:fg:eqf}).

In order to study the asymptotic behavior for the price we must
distinguish between the case $\rho_F\neq 0$ and $\rho_F=0$.


Let us consider first the situation in which $\rho_{F}=0$ (or
equivalently $\rho_C=1$). For this purpose, we introduce the
scaling
$${V}(s,\tau)=\frac{1}{s}{v}(\chi,\tau), \ \ \ \chi = \log(s).$$
It is straightforward to show that ${v}(\chi,\tau)$ satisfies the
following linear convection diffusion equation
$$
\frac{\partial}{\partial \tau}{v}(\chi,\tau) = \left[
\frac{\nu}{2} - \tilde{\beta}Y
t_{C}\right]\frac{\partial}{\partial \chi}{v}(\chi,\tau)
+\frac{\nu}{2}\frac{\partial^{2}}{\partial
\chi^{2}}{v}(\chi,\tau),
$$
which admits the self-similar solution \cite{ch4:CPP} 
$$
{v}(\chi,\tau) = \frac{1}{(2\log(E(\tau)/S(\tau)^2)
\pi)^{\frac{1}{2}}}\exp\left( -\frac{(\chi +
\log(\sqrt{E(\tau)}/S(\tau))-\log(S(\tau)))^{2}}{2\log(E(\tau)/S(\tau)^2)}\right),
$$
with
$$
E(\tau)=\int_0^\infty V(s,\tau) s^2 \,ds.
$$
Then reverting to the original variables it gives the lognormal
behavior \be {V}(s,\tau) = \frac{1}{s(2\log(E(\tau)/S(\tau)^2)
\pi)^{\frac{1}{2}}}\exp\left( -\frac{(
\log(s\sqrt{E(\tau)}/S(\tau)^2)^{2}}{2\log(E(\tau)/S(\tau)^2)}\right),
\label{ch4:eq:logp} \ee where $E(\tau)$ satisfies the
differential equation
\[
\frac{dE}{d\tau}=(2\tilde{\beta}Y t_C+\nu) E(\tau).
\]
Thus for a steady state characterized by $(ii)$ in Proposition
\ref{ch4:prop:MP} we have $S(\tau)=S_0$, $Y=0$ and
$E(\tau)=e^{\nu\tau}E_0$.

Besides the above equilibrium state, equation (\ref{ch4:eq:logp})
characterizes also the self-similar behavior of the price
distribution in the case of booms and crashes, when the price
$S(\tau)$ grows arbitrary or decays to zero. In particular in the
limit $S(\tau)\to 0$, point $(iii)$ in Proposition
\ref{ch4:prop:MP}, the distribution function $V(s,\tau)$
concentrates near zero.

Finally we consider the microscopic behavior of the model where
both $\rho_{C}\neq 0$ and $\rho_{F}\neq 0$.

Recall now the Fokker-Planck equation for the price (\ref{8b}) and
consider the stationary case $(i)$ in Proposition
\ref{ch4:prop:MP}. The Fokker-Planck equation in such case reads
\begin{equation}
\frac{\partial}{\partial \tau}{V} +\frac{\partial}{\partial
s}\left[\tilde{\beta} \rho_F \gamma(S_{F}-s) {V}\right] =
\frac{\nu}{2}\frac{\partial^{2}}{\partial s^{2}}\left(
s^{2}{V}\right).
\end{equation}
In this case the steady state can be computed as \cite{ch4:BM,
ch4:CPT} and yelds
\begin{equation}
{V}^{\infty}(s) =
C_{1}(\mu)\frac{1}{s^{1+\mu}}e^{-\frac{(\mu-1)S_F}{s}},
\label{ch4:eq:sslm}
\end{equation}
where $\mu=1+2\tilde{\beta}\rho_F\gamma /\nu$ and
$C_1(\mu)=((\mu-1)S_F)^\mu/\Gamma(\mu)$ with $\Gamma(\cdot)$ being
the usual Gamma function. Therefore the stationary state is
described by a Gamma-like distribution with Pareto power law
tails.

\begin{remark}~\\
\begin{itemize}
\item
The presence of fundamentalists is then essential in order to
obtain fat tails in the price distribution. Their presence force
the price to approach the mean value $S_F$ in a way similar to the
redistribution of wealth in the models proposed in \cite{ch4:BM,
ch4:CPT}. This feature seems to be essential for the development
of power law behaviors. The stationary state for the price
(\ref{ch4:eq:sslm}) has in fact the same structure of the
stationary states for the wealth in \cite{ch4:BM, ch4:CPT}.
\item
In our description we have considered a constant value for the fundamental price.
Such an assumption might seem quite unrealistic since, according to the economic literature, the fundamental price is usually treated like a
temporal series with a stationary lognormal distribution. This reflect the facts that the returns in logarithmic form are gaussian
distributed with zero mean and a fixed variance, i.e big jumps between two successive realizations are rarely verified.
Note however that introducing a given time dependent distribution function $V_{SF}(q,t)$ for the fundamental price such that
$$S_F(t)=\int_{0}^{+\infty}V_{SF}(q,t)q\,dq,$$
and considering the following dynamic in the price evolution
\[s' = s + \beta (\rho_{C}t_{C}Y(t)s +
\rho_{F}\gamma(q-s)) + \eta s,
\]
where $s$ and $q$ are random variables distributed as $V(s,t)$ and $V_{SF}(q,t)$,
we recover
$$\int_{0}^{\infty}L(V)(s)\varphi(s)\,ds=\int_{0}^{\infty}\int_{0}^{\infty}\int_{\R}
b(s,\eta)V_{SF}(q)V(s)(\varphi(s')-\varphi(s))d\eta\,ds\,dq$$
which is the analogous of (\ref{Priceoperator}) and yields the
same Fokker-Planck equation (\ref{8b}) for the asymptotic behavior
of the model. We omit the details.
\end{itemize}
\end{remark}

\section{Numerical examples}
In this section we considered different numerical simulations of
the kinetic system. A Monte Carlo method analogous to the one used
in classical rarefied gas dynamic has been used for the simulations \cite{ch4:CIP}.
In order to compute the kinetic behavior of
the price, we use a set of $N_s=50000$ samples which can be though
as possible realizations of the random variable $s$ denoting the
price. Since at the initial time the stock price $S_0$ is supposed
to be known, all samples are initialized at the same value
initially. We represent the initial chartists distribution with a set of $N_c=\rho_C(0)N$ sample agents with $N=50000$.
These do not represent real agents but simply statistical realization of the random variable $y$. Such choices of $N_s$ and $N$ permits
to obtain results with a moderate effect of fluctuations without averaging.\\
In all our computations we take the value function
\[
\Phi(x)=
\left\{%
\begin{array}{ll}
\left(\frac{x-R_0}{L-R_0}\right)^r    , & L>x>R_0, \\
-\left(\frac{R_0-x}{R_0+L}\right)^l    , & -L<x \leq R_0, \\
\end{array}%
\right.
\]
where $x\in [-L,L]$, $R_0$ is the reference point and $0<l\leq r <
1$. For example we choose $r=1/2$ and $l=1/4$.

\subsection*{Test 1} In the first test we consider the case with
$\rho_F=0$ i.e only chartists are present in the model. We computed
the equilibrium distribution
for $\Phi(0)=0$ of the investment propensity. We take $\beta=0.1$,
$t_C=1$, a constant herding function $H(y)=1$ and the coefficients
$\alpha_1=\alpha_2=0.01$. The initial data for the chartists is
perfectly symmetric with $Y=0$, so the price remains constant
$S=S_0$ with $S_0=10$. A particular care is required in the
simulation to keep $Y=0$ since the equilibrium point is unstable
and as soon as $Y\neq 0$ the results deviate towards a market boom
or crash.

\begin{figure}
\begin{center}
{\includegraphics[scale=.40]{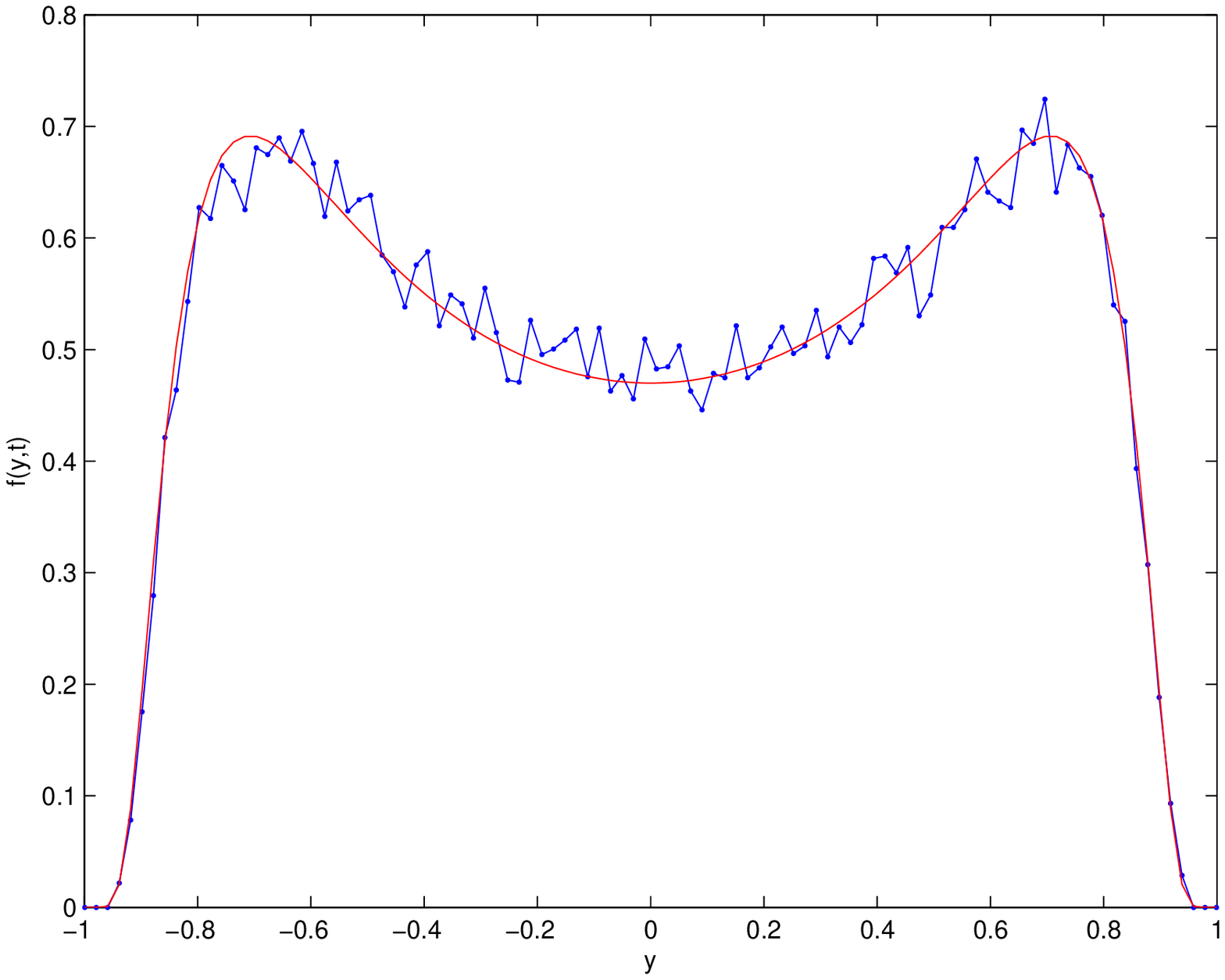}}\,
{\includegraphics[scale=.40]{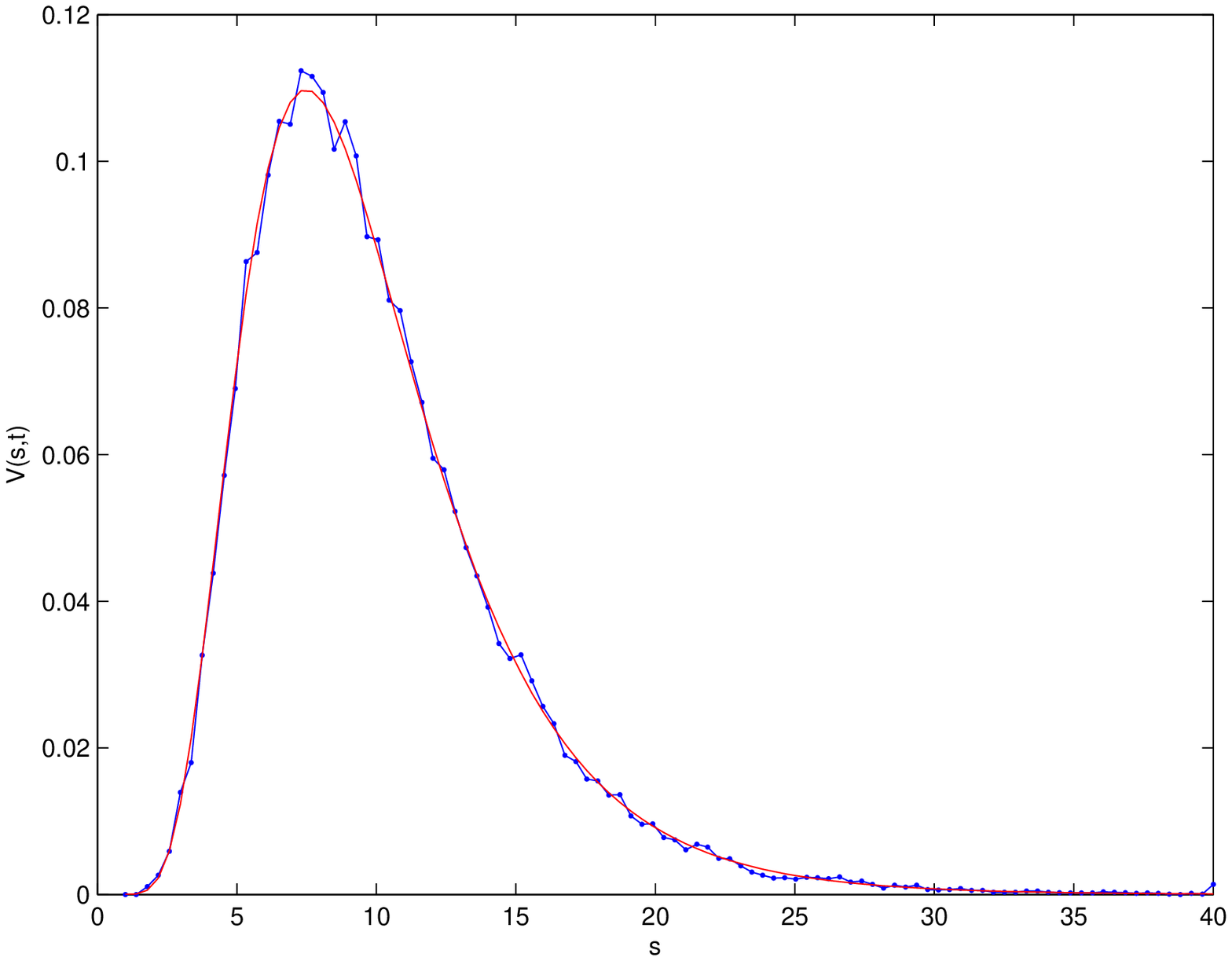}}
\end{center}
\caption{Equilibrium distribution function of the chartist
investment propensity, with $\Phi(0)=0$ (left) and log-normal
distribution for the price (right) at $t=1500$. The continuous
line is the solution of the corresponding Fokker-Planck equation.}
\label{ch4:Mptest1a}
\end{figure}

After $T=1500$ iteration the solution for the investment
propensity has reached a stationary state and is plotted together
with the solution of the Fokker-Planck limit in Figure
\ref{ch4:Mptest1a}. In the same figure we report also the computed
solution for the price distribution and the self-similar lognormal
solution of the corresponding Fokker-Planck equation. A very good agreement between
the computed Boltzmann solution and the Fokker-Planck solution is observed.

\begin{figure}
\begin{center}
{\includegraphics[scale=.40]{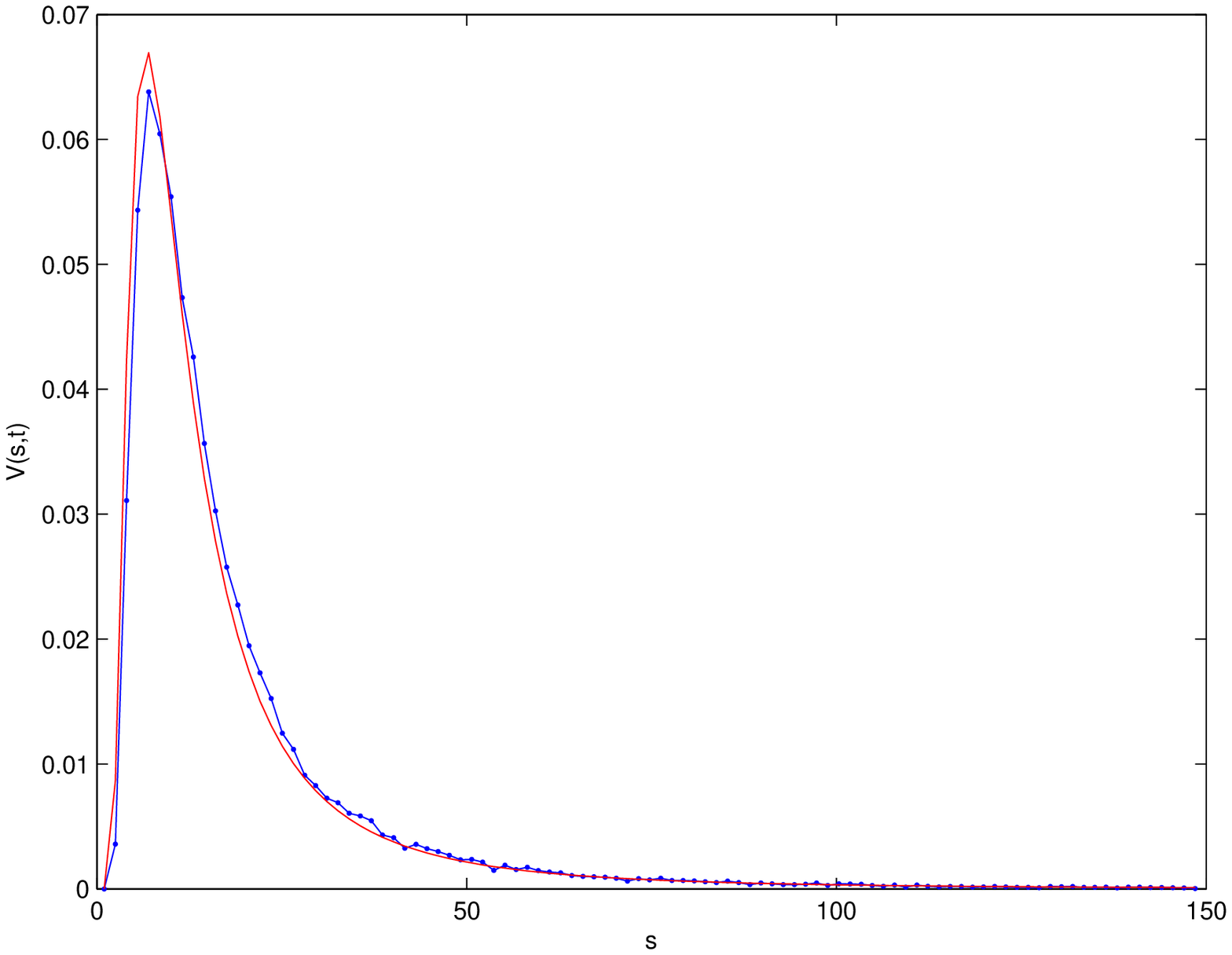}}\,
{\includegraphics[scale=.40]{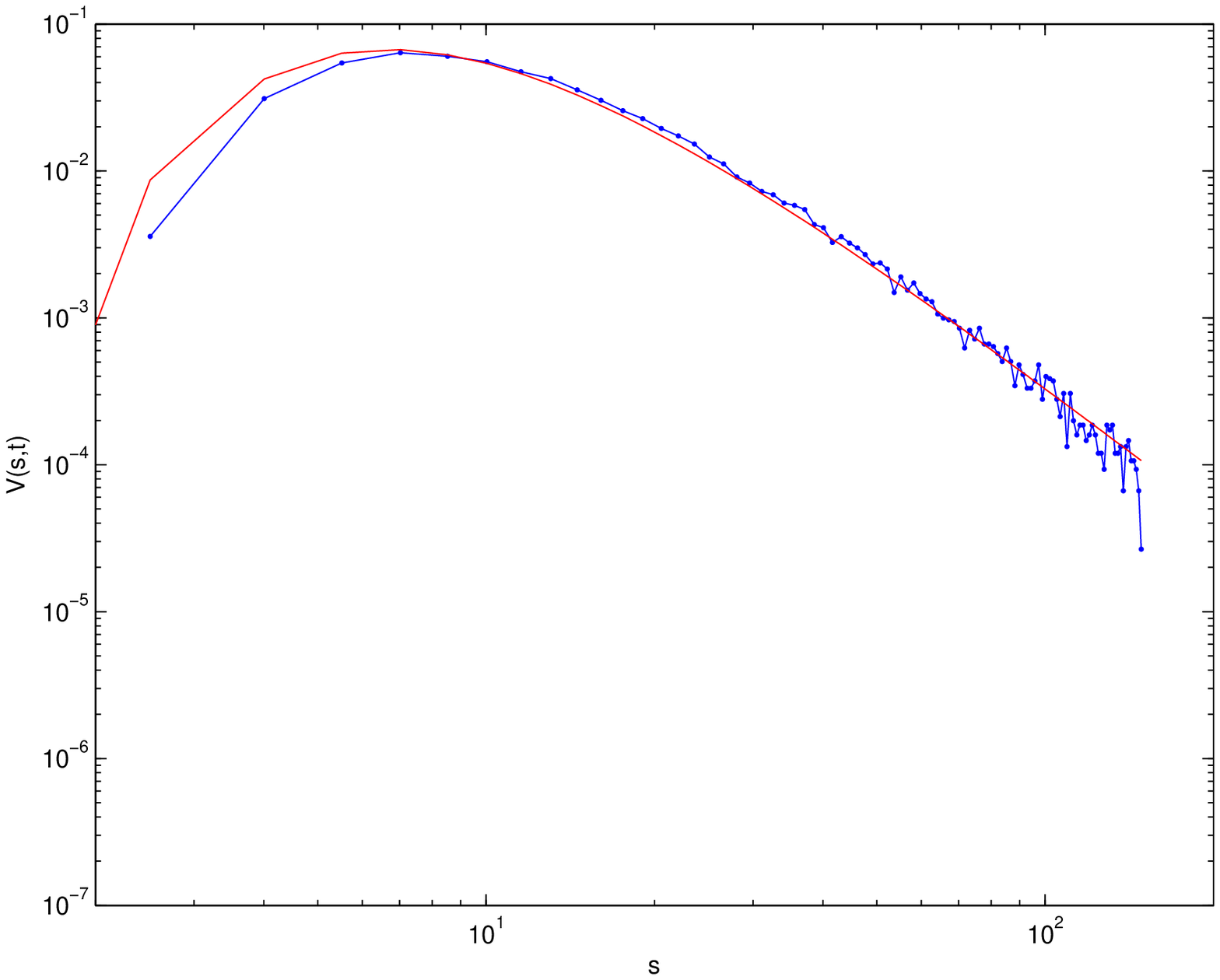}}
\end{center}
\caption{Stationary price distribution for the price with
$\rho_F=\rho_C=0.5$. Figure on the right is in log-log scale. The
continuous line is the solution of the corresponding Fokker-Planck
equation.} \label{ch4:Mptest2}
\end{figure}

\subsection*{Test 2} In the second test case we considered the
most interesting situation with the presence of fundamentalists,
i.e both chartists and fundamentalists interact in the stock
market. We compute an equilibrium situation where $ \rho_F =
\rho_C = 0.5$ and the price stationary at the fundamental value
$S_F=20$. We take
$\beta=0.1$, $t_C=1$, $\gamma=1.3$, $\alpha_1=\alpha_2=0.01$. We
report the result of the simulation for the price distribution at
the stationary state. In Figure \ref{ch4:Mptest2} we show the
price distribution together with the steady state of the
corresponding Fokker-Planck equation. The emergence of a power law
is clear also for the Boltzmann model, and deviations of the two
models is observed for small values of the price.

\subsection*{Test 3}
\begin{figure}[h!]
  \begin{center}
  {\includegraphics[scale=.46]{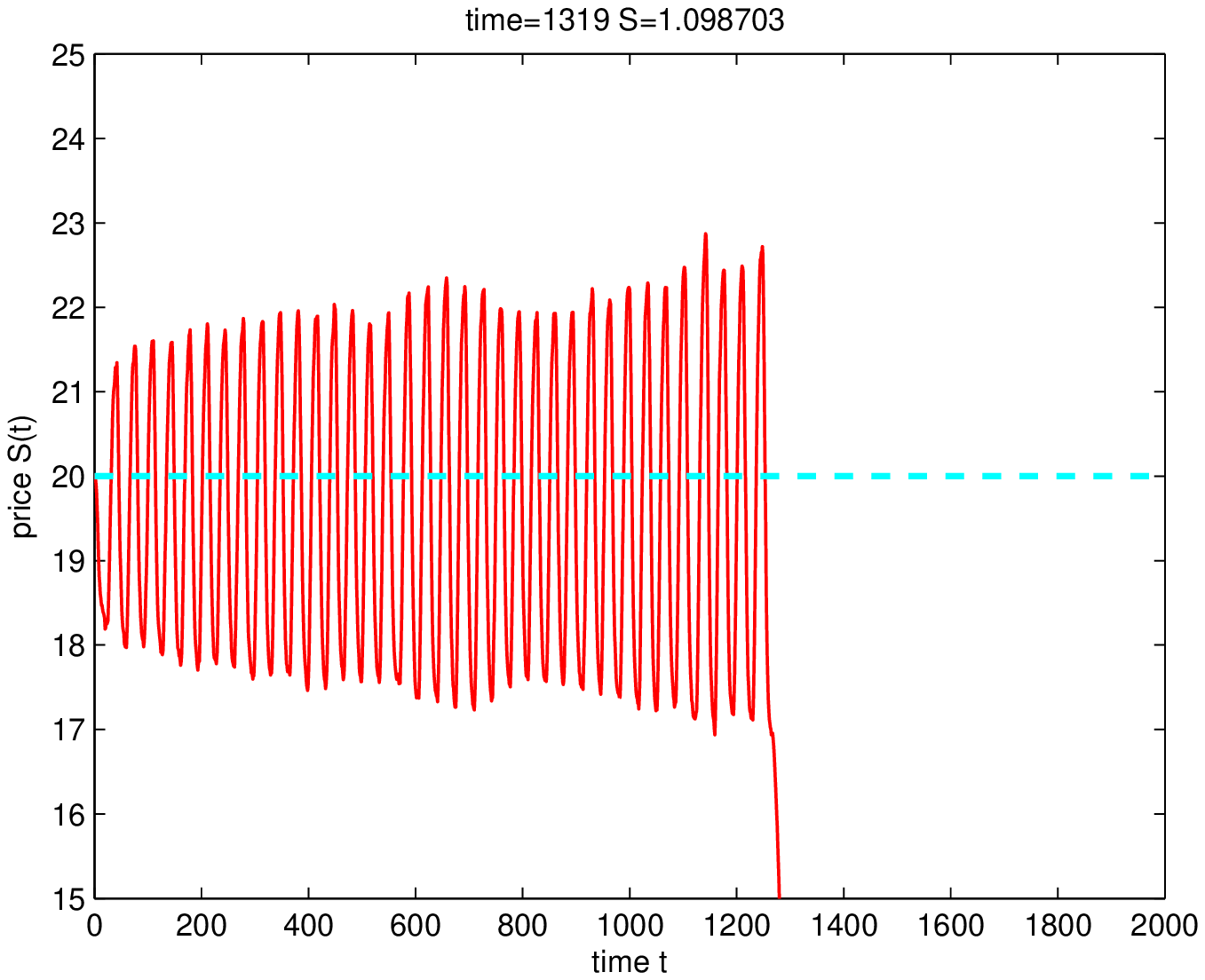}}\,
  {\includegraphics[scale=.46]{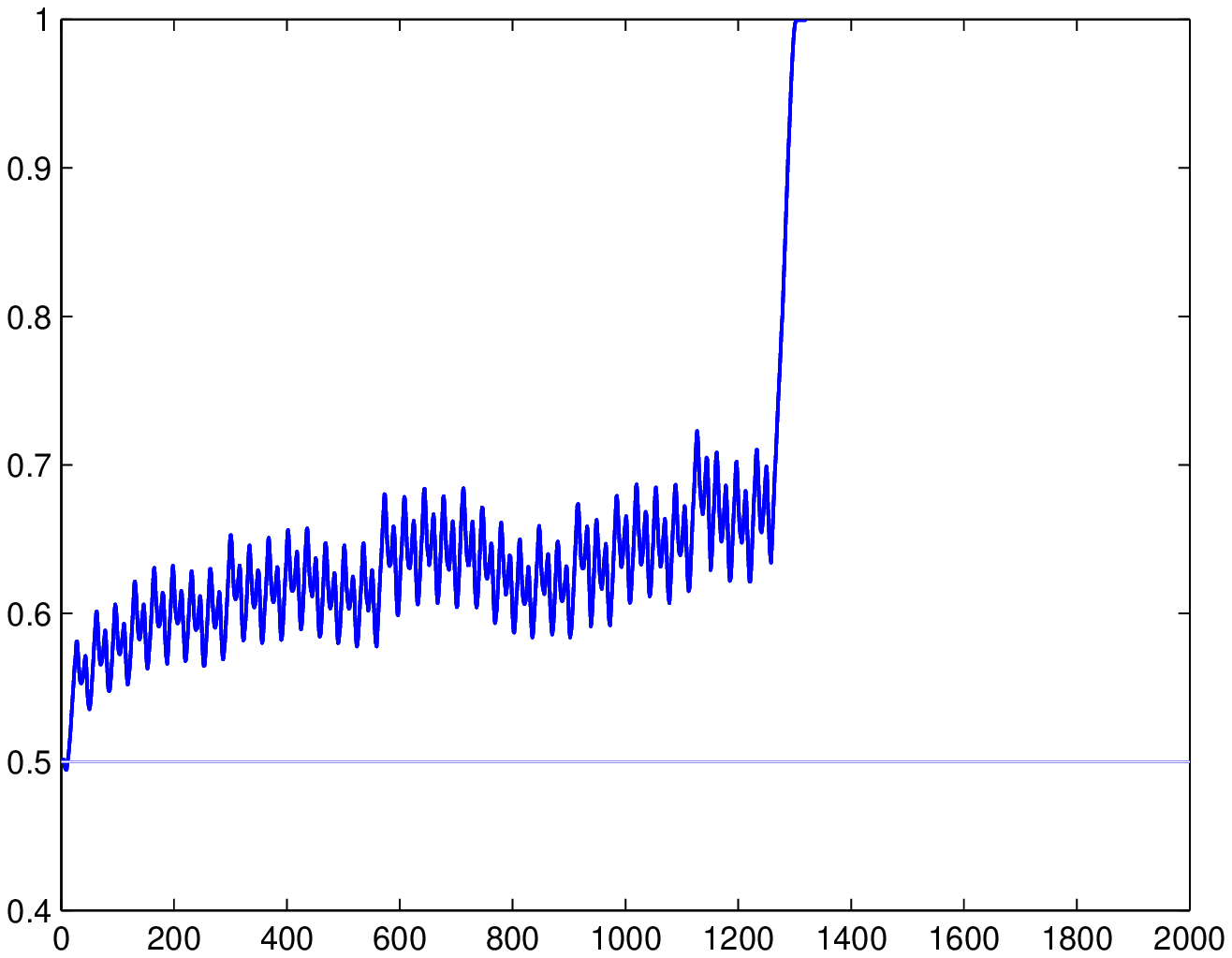}}
  \end{center}
 \caption{Market crash due to a chartist predominance. The plot has been magnified to keep the price scale constant.
 The chartist dynamic is characterized by the parameters
 $\alpha_1=0.2$ and $\alpha_2=0.55$. Figure on the left represent the price averaged over $N_s=50000$ samples.
 Figure on the right represent the variation of the chartists's fraction among the
 entire population of agents.}
 \label{ch4:Mptest1}
\end{figure}

\begin{figure}[h!]
  \begin{center}
  {\includegraphics[scale=.46]{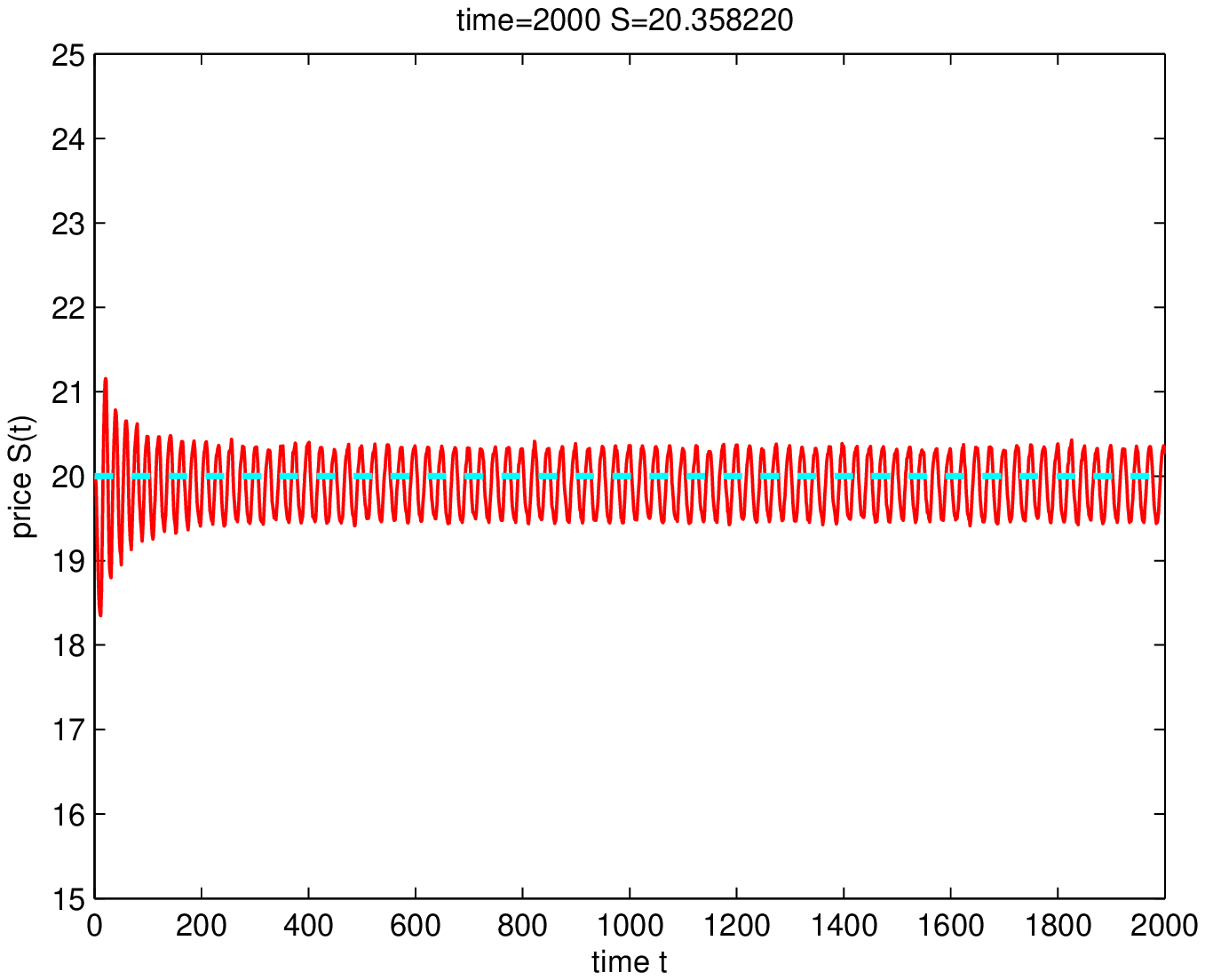}}\,
  {\includegraphics[scale=.46]{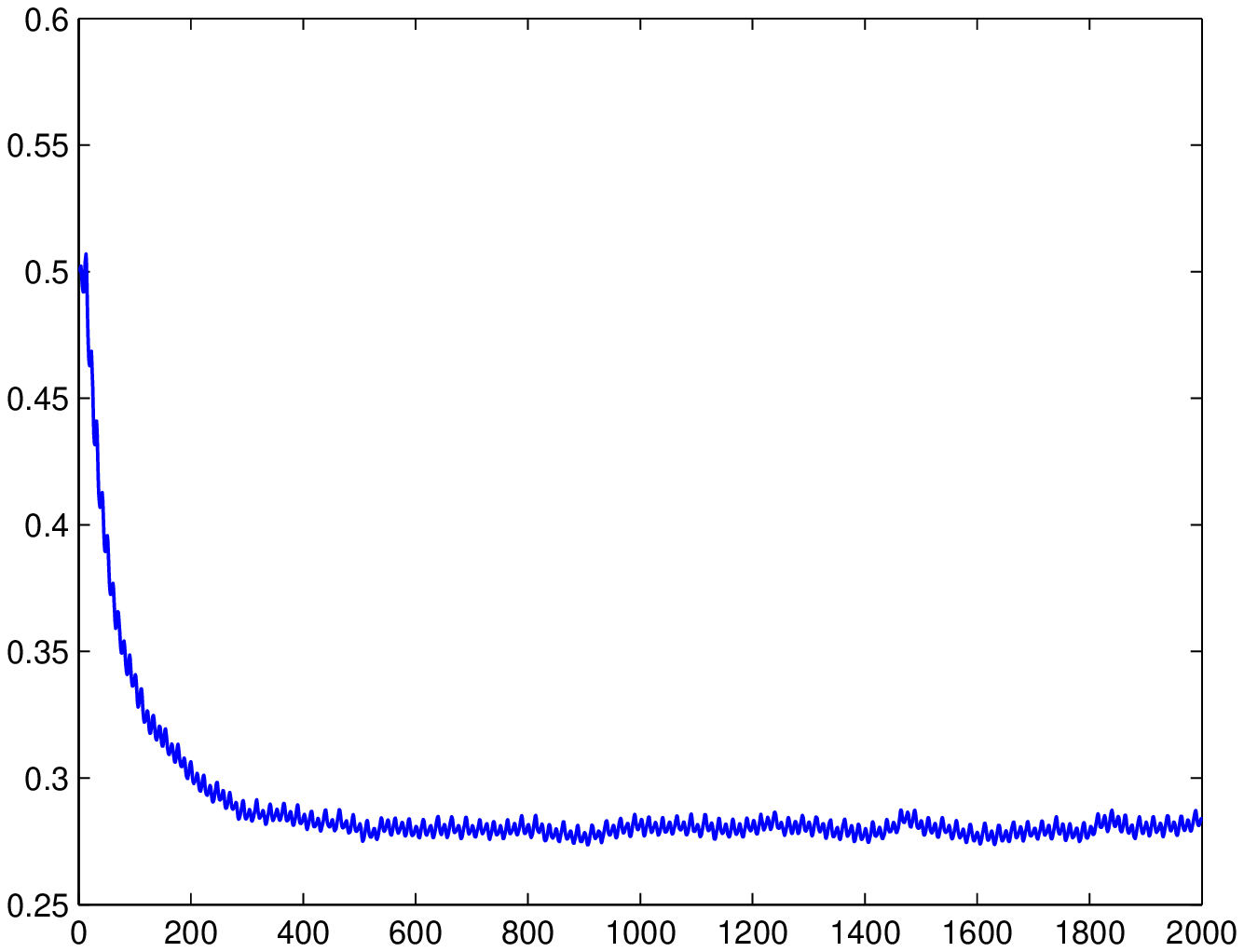}}
  \end{center}
 \caption{Dumped oscillation in the mean price, due to a predominance of fundamentalists. The chartist dynamic is characterized by the parameters
 $\alpha_1=0.2$ and $\alpha_2=0.7$. Figure on the left represent the price averaged over $N_s=50000$ samples. Figure on the right represent the variation of the chartists's fraction among the
 entire population of agents.}
 \label{ch4:Mptest2a}
 \end{figure}

\begin{figure}[h!]
\begin{center}
{\includegraphics[scale=.46]{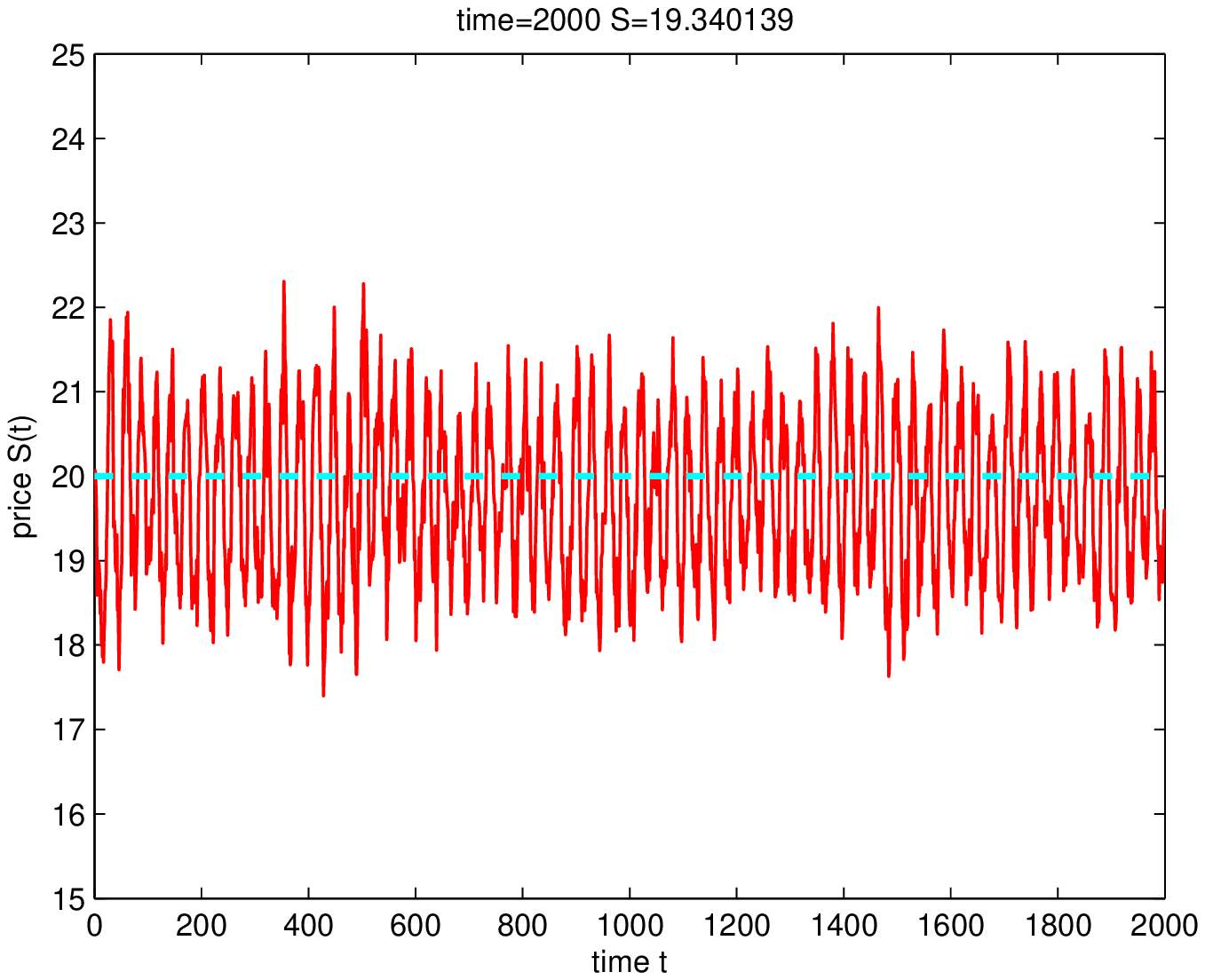}}
{\includegraphics[scale=.46]{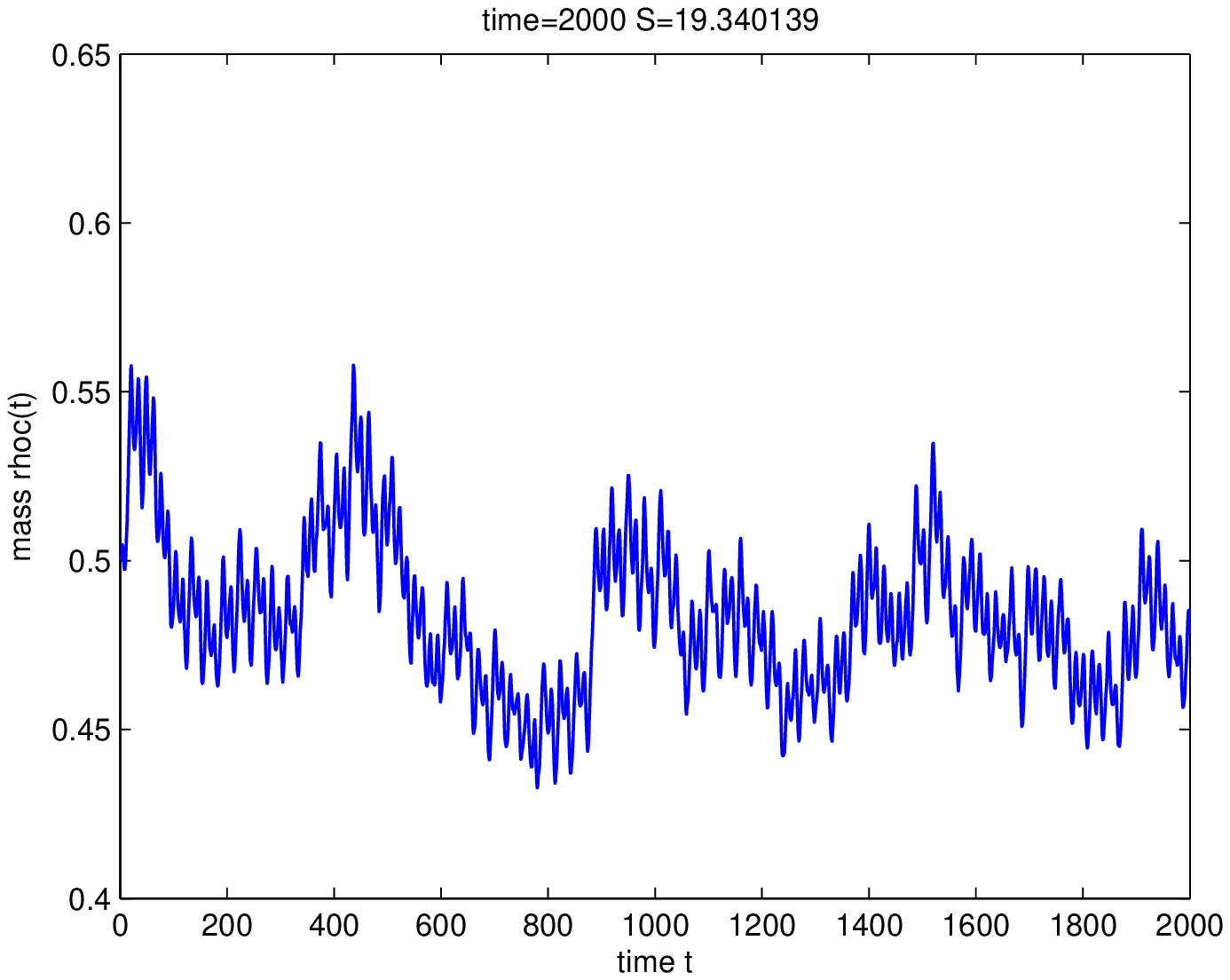}}
 \caption{Oscillations with different amplitudes in the mean  price. The price is computed averaging over $N_s=500$ samples.
 The chartist dynamic is characterized by the parameters
 $\alpha_1=0.5$ and $\alpha_2=0.4$. Figure on the right represent the variation of the chartists's fraction among the
 entire population of agents.}
 \label{ch4:Mptest3a}
\end{center}
\end{figure}

\begin{figure}[h!]
\begin{center}
{\includegraphics[scale=.46]{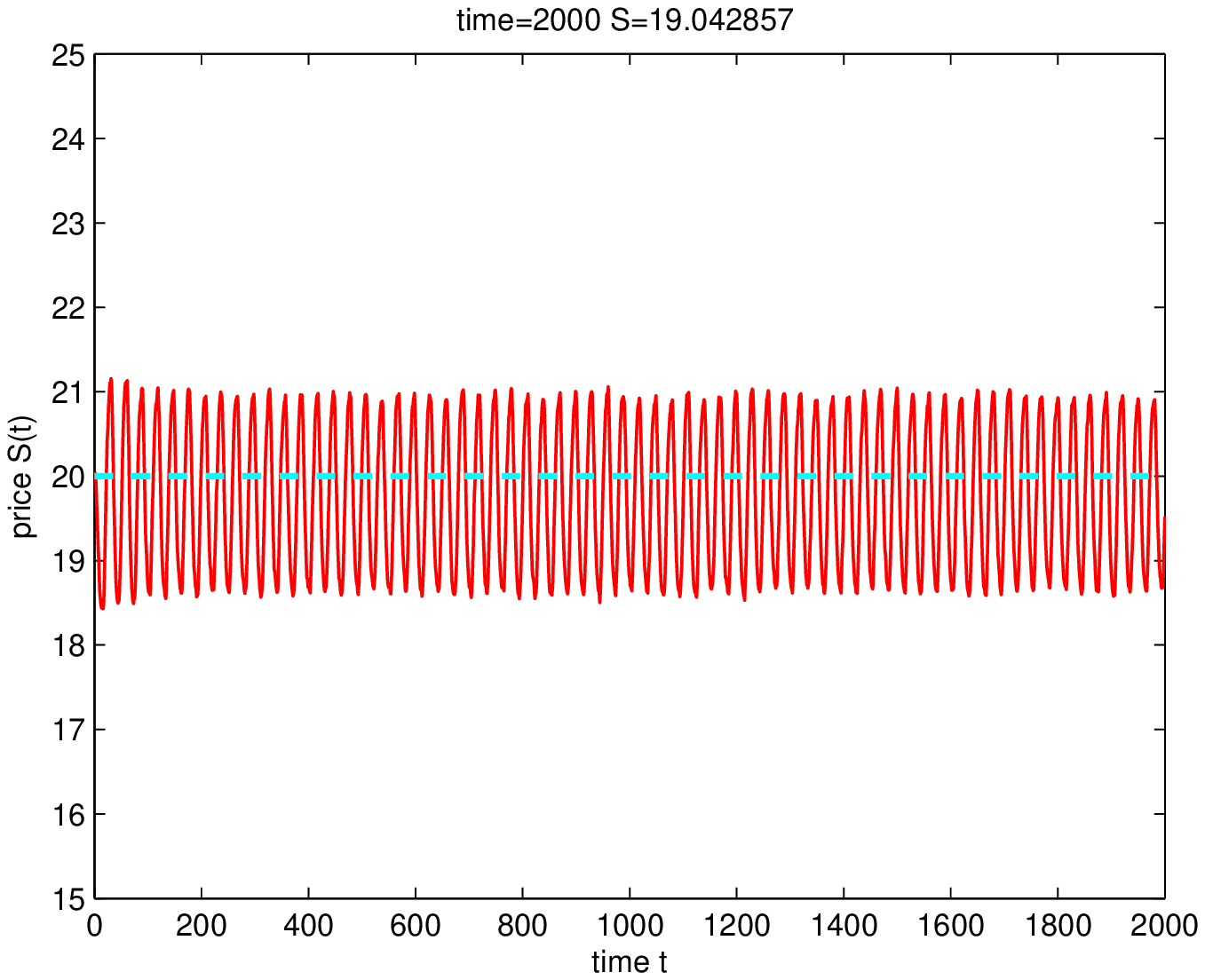}}
{\includegraphics[scale=.46]{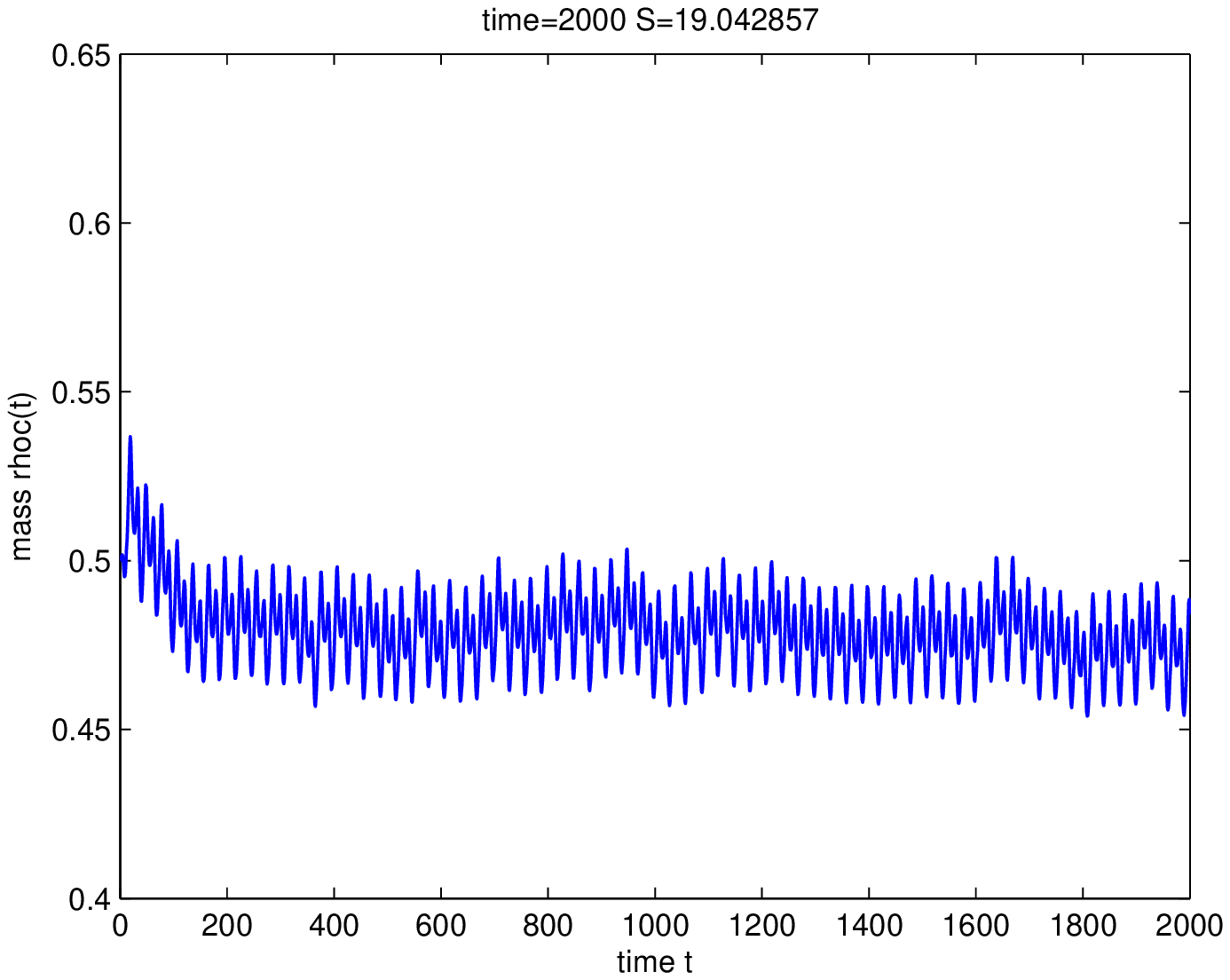}}
 \caption{Same as in Figure \ref{ch4:Mptest3a} but computing the price averaging over $N_s=50000$ samples.}
 \label{ch4:Mptest3c}
\end{center}
\end{figure}

In the third test we consider the case with strategy exchange
between the two populations of interacting agents. The switching
rate used to run the simulation has the following form
$$B_{FC}(x)=e^{\sigma x},$$
where $\sigma$ represent the inertia of the reaction to profit
differentials. We start the simulation considering $\rho_C=\rho_F=0.5$. The fundamental price is
$S_F=20$, we take $\beta=6$, $t_C=0.02$, $\gamma=0.1$,
$\sigma=0.8$, $\mu=0.2$, $D=0.004$, $k=0.75$, and
$\psi(y)=\sgn(y)$. Furthermore we consider an herding function of
the form $H(y)=(1-|y|)$. We run different simulations for $T=2000$
iterations, with different values of $\alpha_1$, and $\alpha_2$,
which measures  respectively the herding and the market influence
on the chartists. Three fundamental behaviors can be observed. The
predominance of chartists, which leads the market towards a crash
or a boom (see Figure \ref{ch4:Mptest1}), the predominance of
fundamentalists, which originates damped oscillation of the price
towards the fundamental value (see Figure \ref{ch4:Mptest2a}), and
a balanced behavior, characterized by periods with
oscillation of the price around the fundamental value
(see Figures \ref{ch4:Mptest3a} and \ref{ch4:Mptest3c}). From
the simulations it is observed that, if we start with a balanced
population between chartists and fundamentalists, the parameter
$\alpha_2$, which characterize the influence of the price trend on
the chartists investment propensity, plays a determinant role in
the competition between the two different trading strategies. In
particular when $\alpha_2\geq 0.6$ fundamentalists are predominant
and price oscillations become dumped.


%

\section{Conclusion}
We derived an interacting agents kinetic model for a simple stock market characterized
by two different market strategies, chartists and fundamentalists.
The kinetic system couples a description for the propensity to invest of chartists and
the price formation mechanism. The model is able to describe several market phenomena like
the presence of of booms, crashes, and cyclic oscillations of the market price. The equilibrium behavior
has been studied in a suitable asymptotic regime which originates a system of Fokker-Planck equation for the chartist's
opinion dynamics and the price formation. We found that in a system of agents acting only using a chartist strategy
the distribution of price converges towards a lognormal distribution. This is in good agreement with what previously
found in \cite{ch4:CPP} and observed in \cite{ch4:LLS1}. When a second strategy based on a fundamentalist approach
is introduced in the model the prices distribution displays Pareto power law tails, which is in accordance to
what observed in the real market data. In the description of the chartists behavior we also introduced a value function which
takes into account the effect of some psychological factors in the opinion formation dynamic. The main effect is
to introduce market instabilities and to reduce the number of stable equilibrium configurations of the system.
Let us finally conclude by observing that in principle several generalizations are possible. We mention here the possibility
to include multiple interacting strategies and/or the influence of the wealth as an independent variable in the
market dynamics.

\appendix
\section{Fokker-Planck asymptotics for the agents distribution}

We report in this appendix the details of the derivation of the
Fokker-Planck equation (\ref{8a}) for the distribution of
chartists. Following \cite{ch4:TG} first we recall the definition
of weak solution for kinetic equations of the form
(\ref{Boltzmanchart}) and (\ref{ch4:eq:price}). Let $I=[-1,1]$ and
$\textit{M}_{p}(I) = \left\{\varTheta\in\textit{M}_{p} :
\int_{I}|y|^{p}d\varTheta(y) < +\infty\right\} $ be the space of
all Borel measure of finite $p$-th order momentum, equipped with
the topology of weak convergence of the measures. Let
$\textit{F}_{s}(I)$ be the class of all real functions $h$ on $I$
such that $h(\pm 1)=h'(\pm 1)=0$
and $h^{(m)}(y)$ is H\"{o}lder continuous of order $\delta$\\
\be \Vert h^{(m)}\Vert_{\delta}=\sup_{y_{1}\neq
y_{2}}\frac{|h^{(m)}(y_{1})-
h^{m}(y_{2})\vert}{|y_{1}-y_{2}|^{\delta}} <\infty \label{eq:hol}
\ee where $0 < \delta \leq 1$, $m+\delta=s$ and $h^{(m)}$ denotes
the $m$-th derivative of $h$.
\begin{definition}
 Let $f^{0}(y)\in \ \textit{M}_{p}(I)$ with $p>1$ an initial probability density, a weak solution for (\ref{Boltzmanchart})
is any probability density $f\in C^{1}(\R^{+},\textit{M}_{p}(I))$ satisfying
\begin{eqnarray}
\frac{d}{dt}\int_{I}f(y,t)\phi(y)dy =
\int_{I^2}\int_{\R^{2}}B(y,y_{*},\eta,\eta_*)f(y)f(y_{*})(\phi(y')-\phi(y))d\eta
d\eta_{*}dy_{*}dy
\end{eqnarray}
for $t>0$ and all $\phi\in\textit{F}_{p}(I)$, and such that
 $$\lim_{t\rightarrow 0}\int_{I}f(y,t)\phi(y) dy = \int_{I}f^{0}(y)\phi(y)dy.$$
\end{definition}
The scaled density $\tilde{f}(y,\tau)$ defined in (\ref{eq:sc1})
satisfies the equation in weak form
\begin{eqnarray}
\frac{d}{d\tau}\int_{I}\tilde{f}(y,\tau)\phi(y)dy =
\frac{1}{\xi}\int_{I^{2}}\int_{J^{2}}\Theta(\eta)\Theta(\eta_{*})\tilde{f}(y)\tilde{f}(y_{*})(\phi(y')-\phi(y))d\eta
d\eta_{*}dy_{*}dy, \label{eq:bw}
\end{eqnarray}
where $J\subseteq \R$ is a suitable symmetric support for the
random variable $\eta$ which avoids the dependence of the kernel
$B$ on the variables $y$ and $y_*$.

Given $\delta \geq 0$ let us take $\phi \in \textit{F}_{2+ \delta}(I)$.\\
From the microscopic dynamic of chartists we have
$$y'-y= \alpha_1 H(y)(y-y_*)+\alpha_2 (\tilde\Phi -y) + D(y)\eta.$$
In the asymptotic limit $\xi\rightarrow 0$, $\sigma^{2}\rightarrow 0$,  we have $y-y'\sim 0$ and
we can use the Taylor expansion
\begin{eqnarray*}
\phi(y')-\phi(y)= \left(\alpha_1 H(y)(y-y_*)+\alpha_2 (\tilde\Phi -y) + D(y)\eta\right)\phi'(y)\\
+\frac{1}{2}\left(\alpha_1 H(y)(y-y_*)+\alpha_2 (\tilde\Phi -y) +
D(y)\eta\right)^{2}\phi''(\tilde{y}),
\end{eqnarray*}
where, for some $0\leq \theta \leq 1$
$$\tilde{y}=\theta y' +(1-\theta)y.$$
Inserting this expansion in the weak formulation of the Boltzman equation, we get
\begin{eqnarray*}
&\displaystyle \frac{d}{d\tau}&\int_{I}\tilde{f}(y,\tau)\phi(y)dy =\\
&\displaystyle
\frac{1}{\xi}&\int_{I^{2}}\int_{J^{2}}\Theta(\eta)\Theta(\eta_{*})\left[\left(\alpha_1
H(y)(y-y_*)+\alpha_2 (\tilde\Phi -y)
+ D(y)\eta\right)\phi'(y)\right.\\
\displaystyle&+&\left.\frac{1}{2}\left(\alpha_1
H(y)(y-y_*)+\alpha_2 (\tilde\Phi -y) + D(y)\eta\right)^{2}
\phi''(y)\right]\tilde{f}(y)\tilde{f}(y_{*})d\eta d\eta_{*}dy_{*}dy \\
&+&R(\xi,\sigma)
\end{eqnarray*}
where
\begin{eqnarray}
\nonumber
R(\xi,\sigma)&=&\frac{1}{2\xi}\int_{I^{2}}\int_{J^{2}}\Theta(\eta)\Theta(\eta_{*})
\left(\alpha_1 H(y)(y-y_*)+\alpha_2 (\tilde\Phi -y) + D(y)\eta\right)^{2}\\[-.2cm]
\label{resto}
\\
\nonumber
&\cdot&(\phi''(\tilde{y})-\phi''(y))\tilde{f}(y)\tilde{f}(y_{*})d\eta_{*}d\eta
dy_{*}dy.
\end{eqnarray}
In order to prove that the remainder (\ref{resto}) goes to zero as $\xi\to 0$ we start observing that, being $\phi\in \textit{F}_{2+\delta}(I)$, and $|\tilde{y}-y|=\theta|y'-y|$ we get
$$|\phi''(\tilde{y})-\phi''(y)|\leq \Vert \phi''\Vert_{\delta}|\tilde{y}-y|^{\delta}\leq \Vert \phi''\Vert_{\delta} |y'-y|^{\delta}.$$
Hence
\begin{eqnarray*}
\displaystyle|R(\xi,\sigma)|&\leq& \frac{\Vert \phi''\Vert_{\delta}}{2\xi}\int_{I^{2}}\int_{J^{2}}\Theta(\eta)\Theta(\eta_{*})\cdot\\
\displaystyle &\cdot& \left|\alpha_1 H(y)(y-y_*)+\alpha_2
(\tilde\Phi -y) + D(y)\eta\right|^{2+\delta}
\tilde{f}(y)\tilde{f}(y_{*})d\eta_{*}d\eta dy_{*}dy.
\end{eqnarray*}
Using the fact that $|H(y)|\leq 1$, $|\tilde{\Phi}|\leq 1$,
$|y|\leq 1$ and applying the following simple inequality
\begin{eqnarray*}
\displaystyle \left|\alpha_1 H(y)(y-y_*)+\alpha_2 (\tilde\Phi -y)
+ D(y)\eta\right|^{2+\delta} \leq
C_\delta\left(\alpha_1^{2+\delta}+\alpha_{2}^{2+\delta}+|\eta|^{2+\delta}\right)
\end{eqnarray*}
with $C_\delta$ a suitable positive constant, \ we finally obtain
$$|R(\xi,\sigma)|\leq C_\delta\rho_C^2\frac{\Vert \phi''\Vert_{\delta}}{2\xi}\left(\alpha_1^{2+\delta}+\alpha_{2}^{2+\delta} +
\int_{J}\Theta(\eta)|\eta|^{2+\delta}d\eta\right).$$ To simplify
computations, we assume that $\Theta$, with zero mean and variance
$\lambda\xi$, is the density of $\sqrt{\lambda\xi}W$ , where $W$
is a random variable with zero mean and unit variance, that
belongs to  $\textit{M}_{2+\alpha}$, for $\alpha>\delta$, so we
have
$$\int_{J}\Theta(\eta)|\eta|^{2+\delta}d\eta = E\left(\left|\sqrt{\lambda\xi}W\right|^{2+\delta}\right) = (\lambda\xi)^{1+\frac{\delta}{2}}E\left(|W|^{2+\delta}\right),$$
and $E\left(|W|^{2+\delta}\right)$ is bounded. This is enough to
show that in the asymptotic limit defined by (\ref{eq:sc2})
the quantity $R(\xi,\sigma)$ tends to zero.\\

Finally taking the limit in the weak formulation yields
\begin{eqnarray*}
\lim_{\xi\to 0}&\displaystyle\frac{1}{\xi}\int_{I^{2}}\int_{J^{2}}&\Theta(\eta)\Theta(\eta_{*})\left[\left(\alpha_1 H(y)(y-y_*)+\alpha_2 (\tilde\Phi -y) + D(y)\eta\right)\phi'(y)\right.\\
&+&\left.\frac{1}{2}\left(\alpha_1 H(y)(y-y_*)+\alpha_2 (\tilde\Phi -y) + D(y)\eta\right)^{2}\phi''(y)\right]\tilde{f}(y)\tilde{f}(y_{*})d\eta d\eta_{*}dy_{*}dy \\
 &=& \int_{I}\left[-\left(\rho_{C}\tilde{\alpha_{1}}
H(y)(Y-y) +
\rho_{C}\tilde{\alpha_{2}}(\tilde\Phi-y)\right)\phi'(y) +
\frac{\lambda}{2}(\rho_{C}{D}^{2}(y))\phi''(y)\right]\tilde{f}(y)dy,
\end{eqnarray*}
which is nothing but the weak form of the Fokker-Planck equation
(\ref{8a}). We can then  state the following theorem

\begin{theorem}
Let the probability density $f^{0} \in {M}_{0}(I)$, and let the
symmetric density $\Theta$ be in $M_{2+\alpha}$ with $\alpha >
\delta$. Then in the asymptotic limit defined by (\ref{eq:sc2})
the weak solution to the Boltzmann equation (\ref{eq:bw}) for the
scaled density $\tilde{f}(y,\tau)$ converges, up to extraction of
a subsequence, to the weak solution of the Fokker-Planck equation
(\ref{8a}).
\end{theorem}

\section{Fokker-Planck asymptotics for the price distribution}

In this appendix we derive the Fokker-Planck limit (\ref{8b}) for
the scaled density distribution of the price. Now let
$\textit{F}_{s}(\R^+)$ be the class of all real functions $h$ on
$\R$ such that $h(0)=h'(0)=0$ and $h^{m}(y)$ is H\"{o}lder
continuous of order $\delta$. We have the following
\begin{definition}
Given an initial price distribution $V_{0}(s) \in
\textit{M}_{p}(\R^{+})$ with $p>1$ a weak solution to
(\ref{ch4:eq:price}) is any probability density
 $V\in C^{1}(\R^{+},\textit{M}_{p}(\R^{+}))$ satisfying
\begin{eqnarray}
\frac{d}{dt}\int_{\R^{+}}V(s,t)\phi(s)ds =
\int_{\R^+}\int_{\R}b(s,\eta)V(s,t)(\phi(s')-\phi(s))d\eta ds
\end{eqnarray}
for $t>0$ and all $\phi\in\textit{F}_{p}(\R^+)$ and such that
 $$\lim_{t\rightarrow 0}\int_{\R^+}V(s,t)\phi(s) ds = \int_{\R^+}\phi(s)V_{0}(s)ds.$$
\end{definition}
Again we start with the weak formulation which now reads
\begin{equation}\label{weakprezzo}
\frac{d}{d\tau}\int_{\R^+}\tilde{V}(s,\tau)\phi(s)ds =
\frac{1}{\xi}\int_{\R^+}\int_{K}\Psi(\eta)\tilde{V}(s)(\phi(s')-\phi(s))d\eta
ds ,
\end{equation}
where $K\subseteq \R$ is a suitable symmetric support for the
random variable $\eta$ which avoids the dependence of the kernel
$b$ on the variable $s$.

Let us take $\phi \in \textit{F}_{2+ \delta}(\R^+)$ with
$\delta>0$. Using a Taylor expansion of $\phi$ around $s$
\begin{eqnarray*}
\phi(s')-\phi(s) &=& \left(\beta(\rho_{C}t_{C}Ys +\rho_{F}\gamma(S_{F}-s))+ \eta s\right)\phi'(s) \\
&+& \frac{1}{2}\left(\beta(\rho_{C}t_{C}Ys
+\rho_{F}\gamma(S_{F}-s))+ \eta s\right)^{2}\phi''(\tilde{s}),
\end{eqnarray*}
where for some $0\leq \theta \leq 1$
$$\tilde{s} = \theta s' + (1-\theta)s, $$
and substituting into (\ref{weakprezzo}) we have
\begin{eqnarray*}
\frac{d}{d\tau}\int_{\R^+}\tilde{V}(s,\tau)\phi(s)ds &=& \frac{1}{\xi}\int_{\R^+}\int_{K}\Psi(\eta)[\left(\beta(\rho_{C}t_{C}Ys +\rho_{F}\gamma(S_{F}-s))+ \eta s\right) \phi'(s)\\
 &+& \frac{1}{2}\left(\beta(\rho_{C}t_{C}Ys +\rho_{F}\gamma(S_{F}-s))+ \eta s\right)^{2} \phi''(s)]\tilde{V}(s) d\eta ds  \\
&+& R(\beta,\zeta,\xi)
\end{eqnarray*}
where
\begin{eqnarray*}
R(\beta,\zeta,\xi) &=&
\frac{1}{2\xi}\int_{\R^+}\int_{K}\Psi(\eta)\left(\beta(\rho_{C}t_{C}Ys
+\rho_{F}\gamma(S_{F}-s))+ \eta s\right)^{2}
\cdot(\phi''(\tilde{s})-\phi''(s))\tilde{V}(s)d\eta ds .
\end{eqnarray*}

Analogously as before, in order to perform the asymptotic limit we
need to show that the quantity $R(\beta,\zeta,\xi)$ approaches
zero as $\xi\to 0$. We observe that being $\phi\in
F_{2+\delta}(\R+)$ and $|\tilde{s}-s|=\theta|s'-s|$ we have
$$|\phi''(\tilde{s})-\phi''(s)|\leq \Vert \phi''\Vert_{\delta}|s'-s|^{\delta}$$
hence
\begin{eqnarray*}
|R(\beta,\zeta,\xi)| \leq \frac{\Vert
\phi''\Vert_{\delta}}{2\xi}\int_{\R^+}\int_{K}
\Psi(\eta)\left|\beta\left(\rho_{C}t_{C}Y
+\rho_{F}\gamma\frac{(S_{F}-s)}{s}\right)+ \eta
\right|^{2+\delta}s^{2+\delta} \tilde{V}(s)d\eta ds .
\end{eqnarray*}
Next we observe that
\begin{eqnarray}
\nonumber
&&\left|\beta\left(\rho_{C}t_{C}Y +\rho_{F}\gamma\frac{(S_{F}-s)}{s}\right)+ \eta
\right|^{2+\delta}\leq\\[-.2cm]
\label{eq:st1}
\\
\nonumber
&&C_{2+\delta}\left((\beta\rho_{C}t_{C})^{2+\delta} +
(\beta\rho_{F}\gamma)^{2+\delta}\left(\frac{S_{F}^{2+\delta}+s^{2+\delta}}{s^{2+\delta}}\right)
+ |\eta|^{2+\delta}\right),
\end{eqnarray}
where $C_{2+\delta}>0$ is a suitable constant.

As in appendix A we assume that $\Psi$, with zero mean and variance $\nu\zeta$ is the density of $\sqrt{\nu\zeta}W$,
where $W$ is a random variable with zero mean and unit variance, that belongs to  $\textit{M}_{2+\alpha}$,
for $\alpha>\delta$, so we have
\begin{equation}\label{etamoment}
\int_{K}\Psi(\eta)|\eta|^{2+\delta}d\eta =
E\left(\left|\sqrt{\nu\zeta}W\right|^{2+\delta}\right) =
(\nu\zeta)^{1+\frac{\delta}{2}}E\left(|W|^{2+\delta}\right),
\end{equation}
and $E\left(|W|^{2+\delta}\right)$ is bounded.

Then we obtain
\begin{eqnarray*}
|R(\beta,\zeta,\xi)| &\leq& C_{2+\delta}\frac{\Vert
\phi''\Vert_{\delta}}{2\xi}\left\{\left[
(\beta\rho_{C}t_{C})^{2+\delta}+(\beta\rho_{F}\gamma)^{2+\delta}+(\nu\zeta)^{1+\frac{\delta}{2}}E\left(|W|^{2+\delta}\right)\right]\right.\\
&\cdot&\left.\int_{\R^+}s^{2+\delta}\tilde{V}(s)ds+
(\beta\rho_{F}\gamma)^{2+\delta}S_{F}^{2+\delta}\right\}.
\end{eqnarray*}
From this inequality it follows that $R(\beta,\zeta,\xi)$ tends to
zero in the limit (\ref{eq:sc3}) if
$$ \displaystyle \int_{\R^+}\tilde{V}(s,\tau)s^{2+\delta}ds$$
is bounded at any fixed time $\tau>0$, provided that the same bound holds at time $\tau=0$ .

To show this we start again from the weak formulation
(\ref{weakprezzo}). The choice $\phi(y)=y^{p}$ gives
$$
\frac{d}{d\tau}\int_{\R^+}\tilde V(s)s^{p}ds =
\frac1{\xi}\int_{\R^+}\int_{K}\Psi(\eta)\tilde
V(s)(s'^{p}-s^{p})d\eta ds .$$ Now
$$s'^{p}-s^{p} = ps^{p-1}(s'-s) + \frac{1}{2}p(p-1)\tilde{s}^{p-2}(s'-s)^{2}$$
where for some $0\leq \theta \leq 1$
$$\tilde{s} = \theta s' + (1-\theta)s.$$
Recalling the microscopic dynamic for the evolution of the price
variable $s$ we can write
\begin{eqnarray*}
\displaystyle \frac{d}{d\tau}\int_{\R^+}\tilde V(s)s^{p}ds &=&
\frac1{\xi}\int_{\R^+}\int_{K}\Psi(\eta)\tilde V(s) \left[
ps^{p-1}(s'-s) +
\frac{1}{2}p(p-1)\tilde{s}^{p-2}(s'-s)^{2}\right]  d\eta ds \\
&=&\displaystyle \frac{p}{\xi}\int_{\R^+}\int_{K}\Psi(\eta)\tilde
V(s)
s^{p-1}\left[\left(\beta(\rho_{C}t_{C}Y s + \rho_{F}\gamma(S_{F}-s)\right) + \eta s\right ] d\eta ds \\
&+&\displaystyle
\frac{p(p-1)}{2\xi}\int_{\R^+}\int_{K}\Psi(\eta)\tilde
V(s)\tilde{s}^{p-2} \left[\beta \left(\rho_{C}t_{C}Y s +
\rho_{F}\gamma{(S_{F}-s)}\right)+ \eta s\right]^{2} d\eta ds.
\end{eqnarray*}
Since the random variable $\eta$ has zero mean value, the first
term in the last expression reduces to
$$p\frac{\beta}{\xi}\left[\left(\rho_{C}t_{C}Y-\rho_{F}\gamma\right)\int_{\R^+}\tilde V(s)s^{p}ds + \rho_{F}S_F\gamma\int_{\R^+}\tilde V(s)s^{p-1}ds\right].$$
For the second therm, we know that
\begin{eqnarray*}
\displaystyle \tilde{s} &=& \theta(s+\beta(\rho_{C}t_{C}Ys +
\rho_{F}\gamma(S_{F}-s)) +\eta s) + (1-\theta)s\\ &=&
\displaystyle s\left[\theta \beta\left(\rho_{C}t_{C}Y +
\rho_{F}\gamma\frac{(S_{F}-s)}{s} \right) +  \theta \eta
+1\right],
\end{eqnarray*}
which implies
\begin{eqnarray*}
\tilde{s}^{p-2} &\leq& {\bar
C}_{p}\left[(\beta\rho_{C}t_{C})^{p-2} +
(\beta\rho_{F}\gamma)^{p-2}\left(\frac{S_F^{p-2}+s^{p-2}}{s^{p-2}}\right)
+ |\eta|^{p-2} + 1\right]s^{p-2},
\end{eqnarray*}
with ${\bar C}_{p}$ a suitable constant.

Gathering all this the weak formulation gives
\begin{eqnarray*}
\displaystyle \frac{d}{d\tau}\int_{\R^+}\tilde V(s)s^{p}ds &\leq&
p\frac{\beta}{\xi}\left[\left(\rho_{C}t_{C}Y-\rho_{F}\gamma\right)\int_{\R^+}\tilde
V(s)s^{p}ds
+ \rho_{F}S_F\gamma\int_{\R^+}\tilde V(s)s^{p-1}ds\right]\\
\displaystyle &+& \frac{p(p-1)}{2\xi}{\bar
C_{p}}\int_{\R^+}\int_{K}\Psi(\eta)\tilde V(s) s^p \left[\beta
\left(\rho_{C}t_{C}Y + \rho_{F}\gamma\frac{(S_{F}-s)}{s}\right)+
\eta\right]^{2}\\
&\cdot& \left[(\beta\rho_{C}t_{C})^{p-2} +
(\beta\rho_{F}\gamma)^{p-2}\left(\frac{S_F^{p-2}+s^{p-2}}{s^{p-2}}\right)
+ |\eta|^{p-2} + 1\right]d\eta ds.
\end{eqnarray*}
Now if we consider the asymptotic limit (\ref{eq:sc3}) and recall
(\ref{etamoment}) for the high order moments of $\eta$, it follows
that the $p$-moments of $\tilde V(s,\tau)$ are bounded at any
finite time independently of $\xi$ and for $p \geq 2+\delta$
satisfy
\begin{eqnarray*}
\displaystyle \frac{d}{d\tau}\int_{\R^+}\tilde V(s)s^{p}ds \leq
A_p\int_{\R^+}V(s,t)s^{p}ds + B_{p}\int_{\R^+}V(s,t)s^{p-1}ds
\end{eqnarray*}
where
$A_p=p\tilde{\beta}\left(\rho_{C}t_{C}Y-\rho_{F}\gamma\right)+{p(p-1)}{\nu\bar
C_{p}}/2$ and $B_{p}=p\tilde{\beta}\rho_{F}S_F\gamma$.\\
Coming
back to the asymptotic expansion we can finally perform the limit
\begin{eqnarray*}
\lim_{\xi \to 0}&\displaystyle\frac{1}{\xi}\int_{\R^+}\int_{K}&\Psi(\eta)\left[\frac{}{}\left(\beta(\rho_{C}(t)Yt_{C}s+\rho_F\gamma(S_F -s)) +\eta s\right)\phi'(s)\right.\\
&+& \left.\frac{1}{2} \left(\beta(\rho_{C}(t)Yt_{C}s+\rho_F\gamma(S_F -s))  +\eta s\right)^{2}\phi''(s)\right]\tilde{V}(s) d\eta ds \\
&=& \int_{\R^+}\left[
\tilde{\beta}(\rho_{C}(t)Yt_{C}s\rho_F\gamma(S_F -s))\phi'(s) +
\frac{\nu}{2}s^{2}\phi''(s)\right]\tilde{V}(s)ds,
\end{eqnarray*}
which is the weak form of the Fokker-Planck equation for the price
(\ref{8b}). So we proved the following
\begin{theorem}\label{fokplankprice}
Let the probability density $V_{0}\in {M}_{0}(\R^{+})$. Then in
the limit defined by (\ref{eq:sc3}) the weak solution to the
Boltzmann equation (\ref{weakprezzo}) for the scaled density
$\tilde{V}(s,\tau)$ converges, up to extraction of a subsequence,
to a weak solution of $(\ref{8b})$.
\end{theorem}

%

\end{document}